\documentclass[12pt, tbtags]{article}

\usepackage{amsfonts}
\usepackage{amssymb, amsmath}
\usepackage{eucal, dsfont, eufrak}
\usepackage{amsthm}
\usepackage{newlfont}
\usepackage[pdftex]{graphicx}
\usepackage{amsopn}

\usepackage{subfig}
\usepackage{ifthen}
\makeatletter
\numberwithin{table}{section}

\newlength{\defbaselineskip}
\@addtoreset{equation}{section}
\def\section{\@startsection {section}{1}{\z@}{-2.5ex plus -1ex minus
-.2ex}{1.3ex plus .2ex}{\large\bf}}
\def\subsection{\@startsection{subsection}{2}{\z@}{-2.25ex plus%
-1ex minus -.2ex}{0.5ex plus .2ex}{\bf}}

\advance \voffset by -0.8in
\advance \hoffset by -0.6in
\def\rel{r}
\def\anj{j}
\def\sig{\zeta}
\def\gren{\varepsilon}
\def\grp{\vec{\pi}}
\def\manif{\mathcal M}
\def\NW{\hat{G}_0}
\def\INW{\hat{G}}
\def\nawi{\hat{\mathfrak{g}}_0}
\def\inawi{\hat{\mathfrak{g}}}

\def\Ad{\mbox{Ad}}
\def\ad{\mbox{ad}}

\def\La{\Lambda}

\def\be{{\mbox{\boldmath $e$}}}

\def\bpm{\begin{pmatrix}}
\def\epm{\end{pmatrix}}

\newcommand{\RR}{\mathbb{R}}
\newcommand{\CC}{\mathbb{C}}

\def\bea{\begin{eqnarray}}
\def\eea{\end{eqnarray}}

\newcommand{\lie}[2]{\langle #1\,,\,#2\rangle}

\newcommand{\dd}{\mathrm{d}}

\newcommand{\g}[1]%
{%
\ifthenelse{\equal{#1}{a}}{\alpha}{}%
\ifthenelse{\equal{#1}{b}}{\beta}{}%
\ifthenelse{\equal{#1}{g}}{\gamma}{}%
\ifthenelse{\equal{#1}{d}}{\delta}{}%
\ifthenelse{\equal{#1}{e}}{\epsilon}{}%
\ifthenelse{\equal{#1}{z}}{\zeta}{}%
\ifthenelse{\equal{#1}{h}}{\eta}{}%
\ifthenelse{\equal{#1}{8}}{\theta}{}%
\ifthenelse{\equal{#1}{i}}{\iota}{}%
\ifthenelse{\equal{#1}{k}}{\kappa}{}%
\ifthenelse{\equal{#1}{l}}{\lambda}{}%
\ifthenelse{\equal{#1}{m}}{\mu}{}%
\ifthenelse{\equal{#1}{n}}{\nu}{}%
\ifthenelse{\equal{#1}{ks}}{\xi}{}%
\ifthenelse{\equal{#1}{p}}{\pi}{}%
\ifthenelse{\equal{#1}{r}}{\rho}{}%
\ifthenelse{\equal{#1}{s}}{\sigma}{}%
\ifthenelse{\equal{#1}{t}}{\tau}{}%
\ifthenelse{\equal{#1}{y}}{\ypsilon}{}%
\ifthenelse{\equal{#1}{f}}{\phi}{}%
\ifthenelse{\equal{#1}{x}}{\chi}{}%
\ifthenelse{\equal{#1}{ps}}{\psi}{}%
\ifthenelse{\equal{#1}{w}}{\omega}{}%
\ifthenelse{\equal{#1}{ff}}{\varphi}{}%
\ifthenelse{\equal{#1}{ee}}{\varepsilon}{}%
}

\begin{document}
\parskip 6pt
\parindent 0pt
\begin{flushright}
EMPG-09-12\\
\end{flushright}

\begin{center}
\baselineskip 24 pt 
{\Large \bf  A Chern-Simons  approach to Galilean quantum gravity in 2+1 dimensions}

\baselineskip 16 pt

\vspace{.5cm}
G.~Papageorgiou{\footnote{\tt georgios@ma.hw.ac.uk} and { B.~J.~Schroers}\footnote{\tt bernd@ma.hw.ac.uk} \\
Department of Mathematics and Maxwell Institute for Mathematical Sciences \\
 Heriot-Watt University \\
Edinburgh EH14 4AS, United Kingdom } \\

\vspace{0.5cm}

{July  2009}

\end{center}

\begin{abstract}
\noindent We define and discuss classical and quantum gravity in 2+1 dimensions in the  Galilean limit. Although there are no Newtonian forces  between massive objects in (2+1)-dimensional gravity, the Galilean limit is not trivial. Depending on the topology of spacetime  there are typically finitely many topological degrees of freedom as well as  topological interactions  of Aharonov-Bohm type between massive objects. In order to capture these topological aspects we consider a two-fold central extension of the Galilei group whose Lie algebra possesses an invariant and non-degenerate inner product. Using this inner product we define Galilean gravity as a Chern-Simons theory of the doubly-extended Galilei group.  The particular extension of the Galilei group we consider  is the classical double  of a much studied group, the extended homogeneous Galilei group,  which  is also  often  called   Nappi-Witten group. We exhibit the Poisson-Lie structure of the doubly extended Galilei group, and quantise the Chern-Simons theory using  a Hamiltonian approach. Many aspects  of the quantum theory are determined by the quantum double of the extended homogenous Galilei group, or Galilei double for short.   We study the representation theory of the Galilei double, explain how associated braid group representations account for the topological interactions in the theory, and briefly comment on an associated non-commutative Galilean spacetime.
\noindent
\end{abstract}

%\centerline{PACS numbers: 04.20.Cv, 02.20.Qs, 02.40.-k}

\section{Introduction}
\label{introsect}

\begin{figure}[htb]
  \centering
 \includegraphics[width =
    0.55\textwidth]{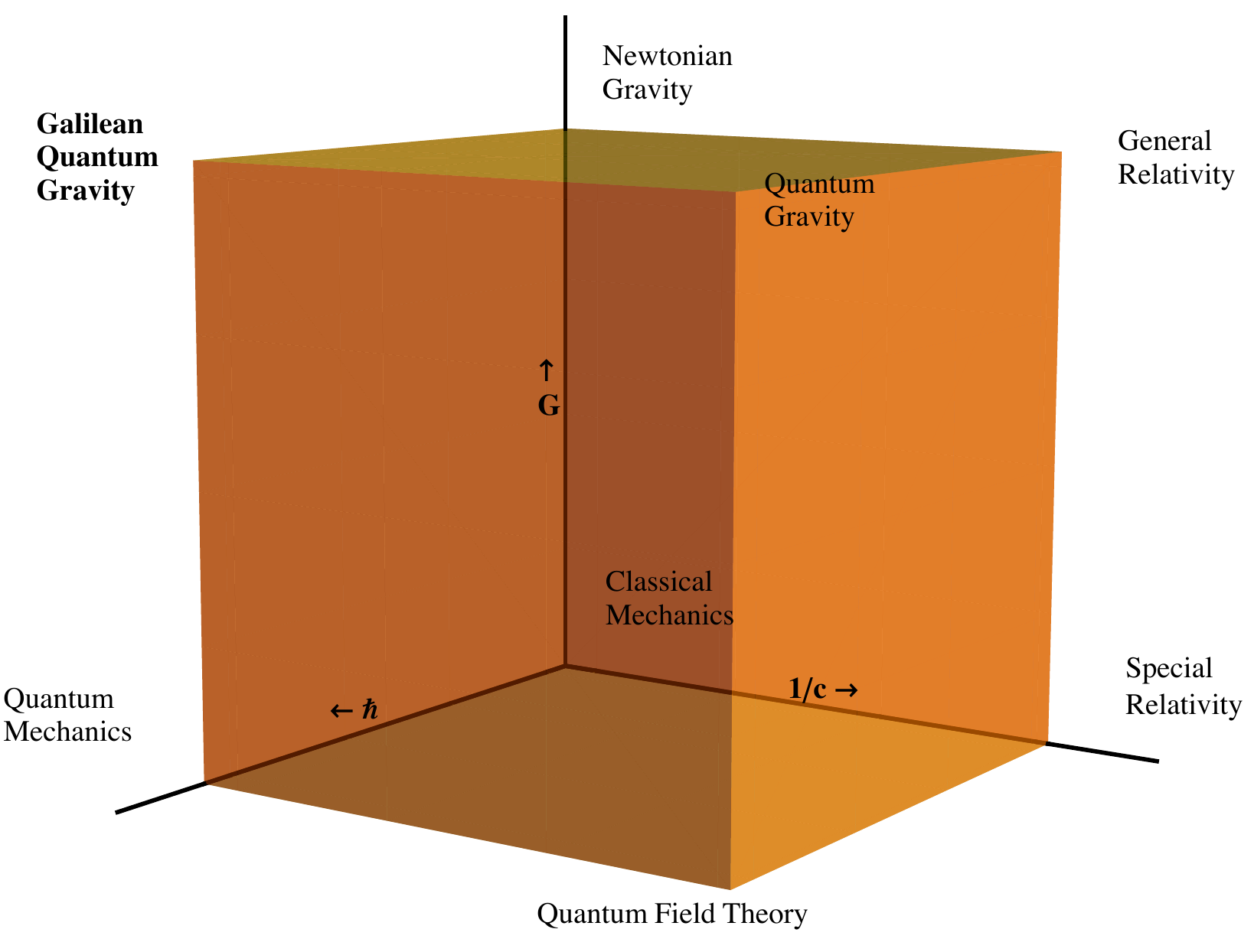}
  \caption{A cube showing  the different regimes of physics,
    based on values of the constants $ c, G$ and $ \hbar $. }
  \label{quantcube}
\end{figure}

The relation between  the different regimes of physics and the values of the fundamental constants 
$c, G$ and $\hbar$  is frequently displayed in  a cube like the one shown  shown in  Figure~\ref{quantcube}. This paper is about the corner of this cube where  $G$ and $\hbar$ have their physical values, but $c=\infty$.  The corresponding physical regime is   non-relativistic or Galilean quantum gravity, which has not received much attention in the literature (but see \cite{Christian} for a comprehensiv recent study and further references). 
 Here we investigate this regime  in a universe with  two space and one time dimension. We construct a diffeomorphism invariant theory of Galilean gravity in this setting,  and show how to quantise it. One motivation for this paper is  to demonstrate that this can be done at all. A related motivation  comes from a curiosity about the dependence of some of the  conceptual and technical problems of quantum gravity on the speed of light. One aspect of this dependence  is that  Galilean gravity  sits between Euclidean and Lorentzian gravity, and  is therefore a good place to study the relation between the two. In the context of three-dimensional gravity, where the relative difficulty of Lorentzian to Euclidean signature is related to the  non-compactness of the Lorentz group versus the compactness of the rotation group, Galilean gravity may provide a  useful intermediate ground, with non-compact but commuting  boost generators.

It is part of the lore of  three-dimensional gravity  that the  Newtonian limit is trivial. If one considers the geodesic equation for a test particle in  2+1 dimensional gravity  and takes the Newtonian limit in the usual way one indeed finds that there are no Newtonian forces on such a test particle \cite{Carlip}.  However, this does not mean that the theory becomes trivial in the  Galilean limit $c\rightarrow \infty$.  Gravity in 2+1 dimensions has topological degrees of freedom associated with the topology of spacetime, and topological interactions between  particles. An example of the latter is a deflection of a test particle in the  metric of a massive particle by an amount which is related to the mass of the latter. Such interactions should survive the Galilean limit.

The question is how to capture topological interactions as $c\rightarrow \infty$. Here we approach this problem from the point of view of the Chern-Simons formulation of 2+1 dimensional gravity.  In this formulation the isometry group of a model spacetime (which depends on the signature and the value of the cosmological constant) is promoted from global  to local symmetry. In the case of Lorentzian signature and vanishing cosmological constant, gravity is thus formulated as a gauge theory of the Poincar\'e group. One might thus attempt to take the Galilean limit by replacing  the Poincar\'e group with the Galilei group.  However, this naive strategy fails since the invariant inner product on the Poincar\'e Lie algebra, which is essential for writing down the Chern-Simons action,  does not have a good  limit as  $c\rightarrow \infty$. In fact, as we shall explain in this paper, the Lie algebra of the Galilei group in 2+1 dimensions does not possess an invariant (and non-degenerate) inner product. 

In order to obtain a group of Galilean symmetries with an invariant inner product on its Lie algebra we consider central extensions of the Galilei group in two spatial dimensions.  Such extensions are ubiquitous in the theoretical description of Galilei-invariant planar systems. One simple reason why central extensions are required to capture all of the physics  when taking the Galilean limit is that the proportionality between rest mass and rest energy of relativistic physics is lost when $c\rightarrow \infty$. Thus, two independent generators are needed to account for mass and rest or 'internal' energy in this limit. In 3+1 dimensions, the Galilei group has only one non-trivial central extension, which is related to the mass.   

In 2+1 dimensions, the Lie algebra of the Galilei group has three distinct  central extensions  \cite{Grigore,Mund,BGGK} and the Galilei group itself  has  two \cite{LevyLeblond,Bose,BGdO}. In the application to non-relativistic particles,  one of the central generators is related to the particle's mass and a   second 'exotic' extension can be  related to the non-commutativity of (suitably defined)  position coordinates of the particle \cite{LSZ,DH0,HP1,HP2} but also to  the particle's spin \cite{BGdO,JN,JN2,NdOT}. There is considerable discussion in the literature regarding this interpretation in general and  in specific models, which we shall not try to summarise here  (but see e.g.  \cite{DH,Hagen}).  In this paper we  adopt the spin interpretation in our notation and terminology. In particular,  in our formalism the spin parameter is naturally paired with the mass parameter,  making manifest an analogy between mass and spin which was stressed in \cite{JN2}. 
Non-commutativity of position coordinates plays a role at the very end of our paper; even though the algebraic structure we obtain is  similar to  the one  discussed in  \cite{LSZ},  the interpretation is different since  our non-commutativity is controlled by universal parameters like $\hbar$ and $G$, and not by parameters characterising an individual particle. The third  central extension  of the Lie algebra does not exponentiate to the Galilei group and  does  not appear to have a clear physical interpretation \cite{BGGK}; it plays no role in this paper. 

   Here we show that  the two-fold central extensions of the Galilei group arises naturally in a framework for taking the Galilean limit which automatically preserves the invariant inner  product on the Lie algebra of the  Poincar\'e group or, more generally, on the isometry group for the relevant value of the cosmological constant.  The invariant product and the associated structures do not appear to have been studied in the literature.  As a by-product of working in a wider context, we obtain central extensions with invariant inner products  of the so-called Newton-Hooke groups \cite{BLL,HS}, which are symmetries of  Galilean spacetimes with a cosmological constant.  Using the inner products on the Lie algebras, one can write down Chern-Simons actions for the extended Galilei and Newton-Hooke groups. In this paper we focus on the case of vanishing cosmological constant. The Chern-Simons  theory for the centrally extended Galilei group with its invariant inner product is our  model for classical  Galilean gravity in 2+1 dimensions. Remarkably, it turns out to be closely related to the gauge formulation of lineal (1+1 dimensional) gravity \cite{CJ}.

    The two-fold central extensions of the Galilei group and the Newton-Hooke groups all have a special property, which  considerably facilitates the  quantisation of the  associated Chern-Simons theory: equipped with their invariant inner product, the Lie algebras of these symmetry groups all have the structure of a classical double. The double structure of the two-fold central extension of the Galilei group has a particularly simple form. The group structure is that of a semi-direct product. The homogeneous part is a central extension of the  homogeneous Galilei group (the group of rotations and Galilean boosts in two spatial dimensions), which has played a role in the context of lineal gravity \cite{CJ} and  in the string theory literature, where it is often called Nappi-Witten group \cite{Nappi:1993ie,FoFS}. The inhomogeneous (normal, abelian)  part is also four-dimensional, and in duality with the Lie algebra of the homogeneous group via the invariant inner product on the Lie algebra.  This is precisely the Lie-algebraic data of a classical double. Moreover, classical doubles also have a Lie co-algebra structure, which can be expressed entirely in terms of a classical $r$-matrix. The $r$-matrix endows the  two-fold extension of the Galilei group (and its Newton-Hooke relatives) with the structure of a Poisson-Lie group. 
 In the context of this paper, the existence of the classical $r$-matrix is important because it allows us to quantise the Chern-Simons formulation of Galilean gravity via the so-called Hamiltonian or  combinatorial quantisation method  \cite{FR,AGSI,AGSII,AS}.

 The starting point of the   combinatorial quantisation method is the construction of the Hopf algebra or quantum group   which quantises the given Poisson-Lie structure defined by the $r$-matrix.  The Hilbert space of the quantum theory as well as  the  action of observables and of large diffeomorphisms  are then constructed in terms of the representation theory of that quantum group.  In the current context, the relevant quantum group is the quantum double of the centrally extended homogeneous Galilei group, which we call the Galilei double. This quantum group is a non-co-commutative deformation of the two-fold extension of the Galilei group. In this paper we do not go through all the details of the combinatorial quantisation programme,  referring the reader instead to the papers listed above and  the papers \cite{Schroers, BNR, BMS, MS1,MS2} discussing its application to three-dimensional gravity.  We do, however, study the Galilei double and its representation theory, and interpret some of its features in terms of Galilean quantum gravity. In particular we show that the non-commutative addition of momenta in Galilean quantum gravity is captured by the representation ring of the quantum  double, and that  distance-independent, topological interactions between  massive particles can be described in terms of a braid-group representation associated to the quantum double. Exploiting a generic link between non-abelian momentum manifolds and non-commutative position coordinates, we briefly comment on the  non-commutative Galilean spacetime associated to the the Galilei double. 
 
The paper is organised as follows. In Sect.~2 we describe a unified
and, to our knowledge, new framework for studying the model spacetimes
of three-dimensional gravity, their isometry groups and the inner
products on the corresponding Lie algebras. We use the language of
Clifford algebras   to obtain the two-fold extension of the
Galilei Lie algebra, with its inner product,  as a contraction limit of a trivial two-fold
extension of the Poincar\'e Lie algebra.   While the contraction procedure for the Lie algebras is well-known \cite{BGdO}, the Clifford language clarifies the origin of the central extensions and naturally provides the inner product.  Similarly, we obtain the
two-fold extensions of the Newton-Hooke Lie algebras as contractions
of trivial central extensions of the de Sitter and anti-de Sitter Lie
algebras. Sect.~3  contains a brief review  of Galilean spacetimes and a
detailed account of  the doubly extended Galilei group. We   explain its relation to the  Nappi-Witten group,  and review  conjugacy classes of the latter. These conjugacy classes, already studied in \cite{FoFS}, play an important role in many calculations in this paper. In particular we show how to deduce coadjoint orbits of the full doubly extended Galilei group from them. 
 These orbits, equipped with a canonical symplectic structure (which we compute), are physically interpreted as phase spaces of non-relativistic particles, possibly with spin.  In Sect.~4 we  introduce the Chern-Simons theory of the doubly extended Galilei group as our model for classical Galilean gravity. We explain how to incorporate point particles by minimal coupling of the Chern-Simons actions to coadjoint orbits of the previous section. The phase space of the theory is the space of flat connection, and can be parametrised in terms of holonomies around non-contractible paths. We  exhibit  the classical $r$-matrix of the doubly extended Galilei group, and briefly explain how this $r$-matrix determines the Poisson structure of the phase space in the formalism of Fock and Rosly \cite{FR}. Finally, Sect.~5 is devoted to the quantisation of Galilean gravity. We explain the role of the  Galilei double  in the quantsisation and 
 study  its representation theory. We show how the associated braid group representation captures topological interactions between massive particles in Galilean quantum gravity, and write down the non-commutative Galilean spacetime  associated to the Galilei double. 

\section{A unified geometrical framework for 3d gravity}

The vacuum solutions of  the Einstein equations in three dimensions are locally isometric to
model spacetimes which depend on the signature of spacetime and on the value of the cosmological constant. As we shall explain below, each of the model spacetimes  can be realised as a hypersurface in a four-dimensional embedding space with a constant metric. It turns out that these four-dimensional geometries and  their Clifford algebras provide a unifying framework for discussing the isometry  groups of the model spacetimes arising in 3d gravity. In particular, the possible invariant bilinear forms on the model spacetimes are naturally associated to the central elements in the even part of the Clifford algebra. For us this framework is valuable because it allows us to take the Galilean limit in such a way that we retain a non-degenerate invariant bilinear form. The resulting version of the Galilei group is naturally a two-fold central extension of the Galilei group in 2+1 dimensions.

We will use Greek indices $\mu,\nu, \ldots $ for the range $\{0,1,2,3\}$ and roman indices $a,b,c, \ldots $  for  $\{0,1,2\}$. Our convention for the spacetime metric in three dimension is   $\eta_{ab}=\text{diag}(-c^2,1,1)$.  We will  mostly be  interested in the Lorentzian case, but the Euclidean case can be included by allowing imaginary  values for $c$. Purely spatial indices will be denoted by Roman letters in the middle of the alphabet $i,j,k...\in\{1,2\}$.  The cosmological constant is denoted $\La$  here (but note that in some papers on 3d gravity, $\Lambda$ stands for minus the cosmological constant in the Lorentzian case \cite{Meusburger, MS6}).

\subsection{Embedding spaces,  Clifford algebras and symmetries}

Consider a  four-dimensional auxiliary  inner product  space $(V,g)$ with metric
\bea
g_{\mu\nu}=\text{diag}(-c^2,1,1,\frac 1 \La).
\eea
We can define model spacetimes as embedded hypersurfaces 
\bea 
\label{hypersur}
H_{c,\La}=
\left\{(t,x,y, w)\in \RR^4| -c^2t^2 + x^2 + y^2 + \frac {1}{\La} w^2 = \frac{1}{\La}\right\} .
\eea
For finite $c$  this gives the three-sphere $S^3$ ($c^2<0$, $ \La >0$), hyperbolic space $H^3$ ($c^2 <0$,  $\La <0$), de Sitter space dS$^3$ ($c^2>0$, $ \La >0$) and anti-de Sitter space AdS$^3$ ($c^2>0$, $\La <0$).
Euclidean  space $E^3$  and Minkowski $M^3$ space arise in the limit  $\La\rightarrow 0$, which one should take {\em after} multiplying the defining equation in \eqref{hypersur} by $\La$. In all cases the  metric is induced by the embedding and has scalar curvature equal to $\La$ (so that $1/\sqrt{|\La|}$ is the curvature radius).

We are interested in the limits $c\rightarrow \infty$ and $\La \rightarrow 0$, or some combination thereof. In those limits, the inverse metric 
\bea
\label{invmet}
g^{\mu\nu}=\text{diag}(-\frac{1}{c^2},1,1,\La)
\eea
degenerates but remains well-defined as a matrix. In studying the associated Clifford algebra (initially for finite $c$ and non-zero $\Lambda$), we therefore work with the inverse metric and define generators $\gamma^a$ via
\bea
\{\gamma^\mu,\gamma^\nu\}=-2g^{\mu\nu}.
\eea
Lie algebra generators of the symmetry group  $SO(V)$ of \eqref{invmet} can be realised as degree two elements in  Clifford algebra
\bea
M^{\mu\nu}=\frac 1 4 [\gamma^\mu,\gamma^\nu],
\eea
giving six generators,  $M^{12}=\frac 1 2 \gamma^1\gamma^2$ etc. with commutation relations
\bea
\label{masterlie}
[ M^{\kappa\lambda}, M^{\mu\nu} ] = 
 g^{\kappa\mu}  M^{\lambda\nu} + g^{\lambda\nu} M^{\kappa\mu}- g^{\kappa\nu} 
 M^{\lambda\mu}-g^{\lambda\mu} M^{\kappa\nu}.
 \eea
 We also write $1$ for the identity element in the Clifford algebra, and define the volume element
\bea
\gamma^5=\gamma^0\gamma^1\gamma^2\gamma^3.
\eea
Then the  linear span of the even  powers  Clifford of the Clifford generators
\bea
{\mathcal A} =\left[1,\gamma^5, M^{\mu\nu}\right]_{\mu,\nu=0,\ldots 3}
\eea  
 is an algebra with multiplication defined by the Clifford multiplication. The elements $1$ and $\gamma^5$ are central elements. Note that 
 the element $\gamma^5 $  satisfies
 \bea
 \label{gamma5norm}
 (\gamma^5)^2 = -\frac{\La}{c^2} =\det(g^{\mu\nu})
 \eea
 and acts on $\mathcal A$ by Clifford multiplication.
  As a Lie algebra with brackets defined by commutators,  $\mathcal A$ is a trivial two-dimensional extension of the  Lie algebra $so(V)$. 
   
 The  elements  $1$ and $\gamma^5$ are, up to scale, the unique elements of, respectively, degree 1 and 4 in ${\mathcal A}$.  Having  chosen them, we can   define maps
 \bea
 \Pi^1, \Pi^5:{\mathcal A}\rightarrow \RR,
 \eea
 which pick out, respectively, the coefficients of $1$ and $\gamma^5$ in the expansion of an arbitrary element in ${\mathcal A}$ in any basis containing $1,\gamma^5$ and otherwise only degree two elements. 
Then we can define   bilinear forms on $\mathcal A$   via
 \bea
 \langle M, N \rangle  = -4\Pi^5(MN),
 \eea
 and 
 \bea
 (M, N) = -4\Pi^1(MN),
\eea
It follows immediately form the centrality of $1$ and $\gamma^5$ that these bilinear forms  are invariant under the conjugation action of  spin groups $Sp(V)$ (which is realised as a subset of $\mathcal A$). Explicitly, the pairing $\langle\;,\;\rangle$
is non-zero whenever the indices on the basis vectors are complementary:
\[
 \langle M^{12},M^{03}\rangle=-1,\quad \langle 1,\gamma^5\rangle =-4,\quad \langle M^{12},M^{01}\rangle=0 \quad \text{etc.}
\]
By contrast,  the pairing $(\;,\;) $  of basis vectors
is non-zero whenever the indices on the basis vectors match:
\[
 (M^{12},M^{12})=1,\quad (M^{01},M^{01})=-\frac{1}{c^2},
 \quad (M^{13},M^{13})=\La \quad \text{etc.}
\]
The  bilinear form  $(\;,\;)$  is the  restriction to even elements of the usual inner product on the Clifford algebra of $(V,g^{\mu\nu})$, induced from the inner product $g^{\mu\nu}$ of degree one elements. Not surprisingly this form  
  degenerates in both  the limits $c\rightarrow \infty$ and $\La\rightarrow 0$. The inner product
  $\langle \;,\;\rangle$, by contrast, 
  remains non-degenerate even when the metric $g^{\mu\nu}$ degenerates.  One can also think of it of the Berezin integral in the Grassmann algebra generated by the $\gamma^\mu$.  The Grassmann algebra is isomorphic to the Clifford algebra as a vector space,
   but clearly independent of the metric $g^{\mu\nu}$.
 The minus signs and factors  in our definition of $\langle \;,\;\rangle$ and $(\;,\;)$ are designed to match conventions elsewhere in the literature. The physical dimensions of the pairings follow directly from the definition: the pairing $\langle \;,\;\rangle$ has dimensions of the inverse of $\gamma^5$ i.e length$^2$/time while the pairing $(\;,\;)$ is dimensionless.

 \subsection{The 2+1 dimensional point of view}
 
 Next we switch notation to one that brings out the spacetime interpretation of the  Lie algebra \eqref{masterlie}.   We define the three-dimensional  totally antisymmetric  tensor with downstairs indices via
 $\epsilon_{012}=1$. Note that 
 \bea
 \label{epsid}
 \epsilon^{abc}=-\frac{1}{c^2}\epsilon_{abc}
  \eea 
  and  that therefore
 \bea
 \epsilon_{abd}\epsilon^{aef}=-\frac{1}{c^2}(\delta^e_b\delta^f_d- \delta^e_d\delta^f_b)
 ,\qquad \epsilon_{abc}\epsilon^{dbc}=-\frac{2}{c^2}\delta_a^d.
 \eea
 Then define
 \bea
 J_a=\frac  1 2 \epsilon_{abc}M^{bc},\quad\quad P^a=M^{a3},
  \eea
  and note that these definitions are independent of the metric $g^{\mu\nu}$. 
  The Lie algebra brackets take the form
    \bea
  \label{21alg}
  [J_a,J_b]=\epsilon_{abc}J^c, \quad [J_a,P^b]=\epsilon_{a\phantom{b}c}^{\phantom{a}b} P^{ c},\quad [P^a,P^b]=-c^2\La
  \epsilon^{abc}J_c.
  \eea
 Because of the identity \eqref{epsid}, the right hand side of the last commutator is actually independent of $c$.  As result, there is no mathematical difficulty in taking the Galilean limit $c\rightarrow \infty$ for the Lie algebra in terms of the generators $J_a$ (indices downstairs) and 
 $P^a$ (indices upstairs).  However, this would not produce the Galilei Lie algebra. To see this,
 we need to study the physical interpretation of the various generators.  First we note the physical dimensions of the generators with indices upstairs and downstairs:
   
  \begin{tabular}{lll}
  $J_0$ :  dimensionless,  &   $J^0$ : 1 /velocity$^2$, & $J_i=J^i$ :  1/velocity \\
  $P_0$:  1/time,   &     $P^0$:  time/length$^2$,   & $ P_i=P^i$ : 1/length.
  \label{dimanal1}
  \end{tabular}

The Lie algebra with generators $J_0, J_1,J_2$  contracts to the algebra of rotations and Galilean boosts in the limit $c\rightarrow \infty$.  However,  since  $\epsilon_{i\phantom{b}j}^{\phantom{a}0}\rightarrow 0
$  as  $c\rightarrow \infty$,  the brackets between $P^0$ and the $J_i$  tends to zero in this limit; if $P^0$ were a time translation generator in the Galilean limit, then this bracket should give a spatial translation. Thus $P^0$ cannot be interpreted as a time translation generator, but $P_0$ can (this is confirmed by  dimensional analysis). Thus, in order to take the Galilean limit we need to write the Lie brackets in terms of $P_a$, with lowered indices.

 The brackets now read
  \bea
  \label{oldconv}
 [J_a,J_b]=\epsilon_{abc}J^c, \quad [J_a,P_b]=\epsilon_{abc} P^c,\quad [P_a,P_b]=-c^2\La
  \epsilon_{abc}J^c,
  \eea
  which is the usual way of writing down the algebra of spacetime symmetry generators in 3d gravity.
    The pairings are
  \bea
  \langle J_a, P_b\rangle =- \eta_{ab} 
  \eea
  or, explicitly, 
  \bea
   \langle J_0,P_0\rangle =c^2,\qquad  \langle J_i,P_j\rangle=-\delta_{ij}
  \eea
  and 
  \bea
  (J_a,J_b)=-\frac{1}{c^2}\eta_{ab},\quad (P_a,P_b)=\La\eta_{ab},
  \eea
which is the same as
\bea
(J_0,J_0)=1, \;(J_i,J_j)=-\frac{1}{c^2}, \; (P_0,P_0)= -c^2\Lambda, \; (P_i,P_j)=\Lambda\delta_{ij}.
\eea
Both these pairings have potential infinities in the limit $c\rightarrow \infty$, and the pairing $(\;,\;)$ has the additional problem of some entries becoming zero.  In the next section we shall see that, by a suitable re-definition of generators  in the bigger algebra $\mathcal A$, the pairing $\langle\;,\;\rangle$ remains well-defined and non-degenerate in the Galilean limit. 
  
  \subsection{The Galilean limit}
  \label{gallimit}
  
 As explained in the introduction,  the difficulties of taking the
 Galilean limit can be  seen easily in terms of  the mass-energy relation. 
  In relativistic physics, mass
 and energy are proportional to each other, but the constant of
 proportionality tends to infinity in the  Galilean limit. Thus, in
 order to retain generators representing mass and energy we expect the
 need for  some kind of renormalisation when taking the speed of light to infinity.

  To make contact with the conventions used in Galilean physics we first rename our generators,
  using the  two-dimensional epsilon symbol  with non-zero entries
  $\epsilon_{12}=-\epsilon_{21} =1$:
  \bea
  \label{KJrel}
\tilde J=-J_0, \qquad  K_i=-\epsilon_{ij}J_j, \qquad 
 \tilde H=-P_0.  
\eea
  The generator $K_i$, $i=1,2$,  generate boosts in the $i$-th spatial direction and we will
  mostly use them rather than the $J_i$ when discussing Galilean physics.
  Then the algebra  \eqref{21alg} reads
  \begin{align}
  \label{explicit}
 [K_i,K_j]&=\epsilon_{ij} \frac{1}{c^2} \tilde J,\quad [K_i, \tilde J]=\epsilon_{ij}K_j 
 \nonumber \\
 [K_i,P_j]&=\delta_{ij}\frac{1}{c^2}\tilde H, \quad [K_i,\tilde H]=P_i,\quad [P_i,\tilde J]=\epsilon_{ij}P_j \nonumber \\
  [P_i,P_j]&=-\epsilon_{ij}\La \tilde J, \quad  [\tilde H,P_i]=-c^2\La K_i .
  \end{align}
  Now we rescale and  rename the central elements  in the algebra ${\mathcal A}$
  \bea
  S=\frac{1}{2c^2}\;1, \quad M=\frac{1}{2}\gamma^5.
  \eea
 As anticipated,  in taking  limit $c\rightarrow \infty$ we need to renormalise the energy $\tilde H$ and the angular momentum $\tilde J$ in such a way that $P^0=\tilde H/c^2$ (the rest mass) and $J^0=\tilde J/c^2$ (a kind of rest spin \cite{JN2})
  remain finite.  We do this by defining
  \bea
   H=\tilde H-c^2M,\quad  J=\tilde J-c^2S.
  \eea
  With this substitution,  the algebra  \eqref{explicit}  
  still  does not have a good limit if we take $c\rightarrow \infty$ while
   keeping $\La$ constant.  The problem is the appearance of $c^2$ in the last line of \eqref{explicit}.     If, on the other hand we let $c\rightarrow \infty$ and $\La\rightarrow 0$ in such a way that $\lambda =-c^2\La$ remains  fixed
we obtain a good limit:
   \begin{align}
   \label{NewtonHooke}
 [K_i,K_j]&=\epsilon_{ij}S\quad [K_i,J]=\epsilon_{ij}K_j \nonumber \\
 [K_i,P_j]&=\delta_{ij} M, \quad [K_i,H]=P_i,\quad [P_i,J]=\epsilon_{ij}P_j \nonumber \\
  [P_i,P_j]&=\epsilon_{ij}\lambda S, \quad [H,P_i]=\lambda K_i .
   \end{align}
   This is a  centrally extended version of the so-called Newton-Hooke algebra, which describes Galilean physics with a cosmological constant $\lambda$ \cite{HS}. 
      
Since our re-definition of generators did not rescale the $J_i$ it is clear that the pairing $(\;,\;)$
degenerates in the limit $c\rightarrow \infty$.  By contrast,  the
pairing $\langle \;,\;\rangle$ remains non-degenerate in the Galilean
limit. Using
\bea
   \langle S,M\rangle =-\frac{1}{c^2},\quad \langle \tilde H,\tilde J \rangle = c^2, 
 \eea 
 we deduce
 \bea
 \label{inprod1}
 \langle H,S\rangle =1,\quad \langle M,J\rangle =1, \quad \langle K_i,P_j\rangle =\epsilon_{ij},
 \eea
 or, equivalently,
   \bea
   \label{inprod2}
\langle H,S\rangle =1,\quad \langle M,J\rangle =1, \quad \langle J_i,P_j\rangle=-\delta_{ij},
\eea
  which is an invariant inner product on the Newton-Hooke algebra, and manifestly independent of  $\lambda$.
The  associated quadratic Casimir, which we will need later in this paper, is
\begin{align}
\label{inwcasimir}
C_2= H\otimes S + S\otimes H + M\otimes J + J\otimes M  - (\vec{J_i} \otimes 
\vec{P_i}+  \vec{P_i} \otimes \vec{J_i}).
\end{align}

Before we proceed it is worth recording the physical dimension of all the quantities introduced so far.  For the generators we have the analogue of \eqref{dimanal1}:

  \begin{tabular}{lll}
  $J$ :  dimensionless,  &   $S$ : 1 /velocity$^2$, & $J_i=J^i$ :  1/velocity \\
  $H$:  1/time,   &     $M$:  time/length$^2$,   & $ P_i=P^i$ : 1/length.
  \label{dimanal2}
  \end{tabular}

  It follows that the pairing $\langle\;,\;\rangle$ is also dimensionful with  dimension length$^2$/time. Finally, the dimension of $\lambda $ is 1/time$^2$, so that it parametrises some kind of curvature in time.

   The smallest  Lie subalgebra   of  \eqref{NewtonHooke}  containing boosts and rotations is the Lie algebra  with generators $J,K_1,K_2,S$ and brackets
      \bea
   [K_1,K_2]&= S\quad [K_i,J]=\epsilon_{ij}K_j  
   \eea
   or, in terms of $J_i=\epsilon_{ij}K_j$, 
       \bea
    \label{nawij}
   [J_1,J_2]&= S\quad [J_i,J]=\epsilon_{ij}J_j.  
   \eea
   This Lie algebra can be viewed as a central extension of the Lie algebra of the Euclidean group in two dimensions,  or as the Heisenberg algebra with an outer automorphism $J$. In  the string theory literature it is sometimes called  the Nappi-Witten Lie algebra \cite{Nappi:1993ie,FoFS}. 
 In the context of this paper it should be interpreted as  central extension of the homogeneous part of the Galilei Lie algebra, so we denote it by   $\nawi$.
  This algebra also has an invariant inner product (of Lorentzian signature!) with non-zero pairings
in the above basis given by 
\bea
\label{NWinprod}
\langle J,S\rangle_{\nawi}=1, \quad \langle J_i,J_j\rangle _{\nawi}=-\delta_{ij}.
\eea
Note that we also have $\langle K_i,K_j\rangle_{\nawi}=-\delta_{ij}$.

One can view the Lie algebra  \eqref{NewtonHooke} as a  'generalised complexification'
 of the homogeneous  algebra $\nawi$. This viewpoint has proved useful in the context of 
 (relativistic) three-dimensional gravity, where it was introduced for vanishing cosmological constant  in \cite{Martin}, generalised to arbitrary values of the  cosmological  constant in  \cite{Meusburger} and further developed in \cite{MS6}.  We will often adopt it in the current paper since it provides the most efficient method for explicit calculations in many cases. The idea, explained in detail  in \cite{Meusburger},   is to work with a formal parameter $\theta$ that satisfies $\theta^2=\lambda$ and to work over the ring of  numbers of the form $a+\theta b$, with $a,b\in\RR$.  The formal parameter $\theta$ is a generalisation of the imaginary unit $i$ since it square may be negative (ordinary complex numbers), positive (hyperbolic numbers) or zero (dual numbers).   In this formulation, we recover \eqref{NewtonHooke} from \eqref{nawij} by setting $H=\theta J, P_i=\theta J_i$ and $M=\theta S$. The  inner product $\langle \;,\;\rangle$  \eqref{inprod1} is then  the 'imaginary part' (linear in $\theta$)  of the linearly extended inner product $\langle\;,\;\rangle_{\nawi}$ \eqref{NWinprod}.

  In this paper we will concentrate on the limiting situation where $\lambda=0$.
  In this limit we obtain a  two-fold extension of the  Galilei Lie  algebra, which we denote $\inawi$:
      \begin{align}
    \label{extgal}
 [K_i,K_j]&=\epsilon_{ij}S\quad [K_i,J]=\epsilon_{ij}K_j \nonumber \\
 [K_i,P_j]&=\delta_{ij} M, \quad [K_i,H]=P_i,\quad [P_i,J]=\epsilon_{ij}P_j \nonumber \\
  [P_i,P_j]&=0, \quad [H,P_i]=0.
   \end{align}
  The associated  two-fold  central extension of the Galilei group, its coadjoint orbits and irreducible representations (irreps)  have been studied extensively \cite{Mund,Grigore,BGdO,NdOT}.
 In the next section we need to revisit some of those results in a 
   language that, amongst others,  makes use of the formal parameter $\theta$ and  of the inner product \eqref{inprod1}.  This language is designed to ease the deformation of the extended Galilei group to the Galilei quantum double in Sect.~\ref{quantumsection}.

\section{The centrally extended Galilei group in two dimensions, its Lie algebra and coadjoint orbits}
\label{galgroupnorbits}

\subsection{Galilean spacetime and symmetry}
\label{galfoundations}
The geometrical  structure of Galilean spacetimes is discussed in a number of textbooks. We base our discussion on the treatment in 
 \cite{Arnold}.  Thus  a  (2+1)-dimensional Galilean spacetime is a three-dimensional affine space $A^3$, with an affine map
 \bea
 t: A^3 \rightarrow \RR,
 \eea
 which assigns to every event an absolute time. After a choice of origin we can identify $A^3$
 with a three-dimensional vector space $V^3$.  With the standard choice of origin for $\RR$, the map $t$ then defines a linear map $V^3\rightarrow \RR$, which we denote by $t$ again. The kernel of this map is the space of simultaneous events. It is a two-dimensional Euclidean vector space. We will pick a basis for $V^3$ consisting of a vector $\be_0$  which is normalised  via $t(\be_0)=1$ and orthonormal vectors $\be_1, \be_2$ in the kernel of $t$. Then we can describe events in terms of coordinates $t,x_1,x_2$
 relative to this basis, and identify $V^3$ with $\RR^3$ (we have re-interpreted the map $t$ as a coordinate here, but this should not lead to confusion). 

In discussing  $\RR^2$ with its standard metric we write  vectors  as  $\vec{a},\vec{b}\ldots $ and  $\epsilon$ for the matrix with elements $\epsilon_{ij}$ i.e.
  \bea
  \epsilon = \bpm 0  &  1 \\
   -1  &  0 \epm.
\eea
We use  the Einstein summation convention  for the trivial metric $\delta_{ij}$ and  
introduce the  abbreviations
\bea
\vec{a}\times \vec{b}= a_i\epsilon_{ij} b_j, \qquad \vec{a}\cdot\vec{b} =a_i b_i \qquad
\vec{a}^2 =\vec{a}\cdot\vec{a}.
\eea

The orthochronous Galilei group is, by definition, the group of affine  maps $A^3\rightarrow A^3$ which preserves   $t$ and the Euclidean structure on  the space of simultaneous events. We will mainly be interested in the identity component  $G_+^{\uparrow}$ of the 
 orthochronous Galilei group   and  its universal cover
 $ \tilde{G}_+^{\uparrow} $, where we used the notation from \cite{Grigore}. 
 Then $G_+^{\uparrow}$  consists of time- and space translations, spatial rotations and Galilean boosts.  Parametrising time- and space translation in terms of $(\g{a},\vec{a})\in \RR^3$,
 rotations in terms of $\varphi\in [0,2\pi)$ and boosts in terms of velocity vectors $\vec{v}\in \RR^2$, we identify $G_+^{\uparrow} =(S^1 \ltimes \RR^2)\ltimes \RR^3 \ $,  write elements in the  form $(\varphi,  \vec{v};\g{a},\vec{a})$ and obtain the  group multiplication law
\begin{align}
&(\varphi_1 ,\vec{v}_1; \g{a}_1, \vec{a}_1 ) (\varphi_2,\vec{v}_2; \g{a}_2, \vec{a}_2) \nonumber \\
&= (\varphi_1 + \varphi_2, R(\varphi_1) \vec{v}_2 +
\vec{v}_1;
  \alpha_1 + \alpha_2, \vec{a}_1 + R(\varphi_1) \vec{a}_2 + \g{a}_2 \vec{v}_1)
\end{align}
with 
\begin{equation}
  \label{eq:1}
  R(\varphi) = \left(\begin{matrix} 
 \cos{\varphi}& \sin{\varphi}  \\ 
- \sin{\varphi} & \cos{\varphi}
\end{matrix}\right).
\end{equation}\\
The action of the Galilei group on $A^3$ can be expressed in terms of  the coordinate  vector
 $ (t,\vec{x}) \in \mathbb{R}^3 $ as follows:
\begin{equation}
(\varphi,  \vec{v};\g{a},\vec{a}): (t, \vec{x})\mapsto ( t +\g{a}, R(\varphi)\vec{x}+\vec{v}t +
\vec{a}).
\end{equation}

The Lie algebra $ \mathfrak{g} $ of $G_+^{\uparrow} $  is conventionally described in terms of  a rotation generator 
$ J $,  boost generators $ K_1,K_2$ and time- and space translation generators $ H, P_1,P_2$.
The non-vanishing brackets are
\bea
\label{galilei_nonext}
 [K_i,H]=P_i \qquad  [P_i,J]=\epsilon_{ij}P_j \qquad  [K_i,J]=\epsilon_{ij}K_j.
\eea
It is easy to check that $P_i\otimes P_i$ and $\epsilon_{ij}(P_i\otimes K_j - K_i\otimes P_j)$ are both central elements in the  universal enveloping algebra  $U(\mathfrak{g})$ and that no linear combination of them is invertible i.e. associated to an inner product on $\mathfrak{g}$.
 Since the dimension of the centre of $U(\mathfrak{g})$ is bounded by the rank of $\mathfrak{g}$, we conclude that the Lie algebra $\mathfrak{g}$ cannot have an invariant, non-degenerate inner product. If it did, the associated Casimir would provide a third element of the centre of $U(\mathfrak{g})$, contradicting rk$(\mathfrak{g})=2$.
 
As reviewed in the Introduction, the Galilei algebra \eqref{galilei_nonext}, admits three central
extensions - one for each two-cocycle in the Lie algebra cohomology \cite{Mund,Grigore}.  The two-fold central extension $\inawi$ \eqref{extgal} we constructed via 
contraction in the previous  section  is  the maximal extension of the Lie algebra that can be exponentiated   to the Galilei group. We now introduce a formulation of this extension as a classical double which is particularly suitable for our purposes.
 
\subsection{The centrally extended homogeneous  Galilei group}   
  
  The  centrally extended  homogeneous Galilei group $\NW$  is a Lie group whose Lie algebra is the  Lie algebra $\nawi$. As a manifold it is  $S^1\times \RR^3$.  Group elements can be parametrised  in
  terms of tuples
  \bea
  \label{nawipara}
(\varphi,  \vec{w},\sig), \qquad \varphi \in [0,2\pi), \; \vec{w}\in \RR^2, \sig \in \RR,
\eea
 which we  we can identify with  exponentials  of the abstract generators via
  \bea
  (\varphi,\vec{w},\sig ) \leftrightarrow  \exp(\vec{w} \cdot \vec{J})\exp(\sig S)\exp(\varphi J).
  \eea

The group composition law is given in 
\cite{FoFS} in complex notation and with slightly different conventions from ours (their generator $J$ is the negative of ours, and the inner product is also the negative of ours).
We choose conventions which allow us to make contact with  the papers \cite{Mund}
and \cite{Grigore} on the Galilean group and its central extension.
Using vector  notation,  the composition law  can be written as 
   \bea
   \label{nawimulti}
   (\varphi_1,\vec{w}_1,\sig_1) (\varphi_2,\vec{w}_2,\sig_2)=(\varphi_1+\varphi_2,
   \vec{w}_1 + R(\varphi_1)\vec{w}_2,\sig),
   \eea
   where $ R(\varphi) $ is an $ SO(2) $ matrix given in \eqref{eq:1}
   and 
   \bea
   \label{sigmult}
   \sig = \sig_1+\sig_2 +\frac 1 2 \vec{w}_1\times R(\varphi_1)\vec{w}_2. 
   \eea 
   The inverse is
 \bea
  (\varphi,\vec{w},\sig )^{-1}= (-\varphi,-R^{-1}(\varphi) \vec{w}, - \sig ).
 \eea
The formula for  group conjugation in the  group $\NW$  plays a fundamental role in what follows.  By explicit calculation, or by translating the result of \cite{FoFS} into our  notation,  one finds, in terms of
   \bea
   \label{vnotation}
 v = (\varphi, \vec{w}, \sig), \qquad  v_0=(\varphi_0, \vec{w}_0, \sig_0),
 \eea
 that 
 \begin{align}
 \label{ccformula}
\Ad_v (v_0)= 
\bigl(\varphi_0, (1- R(\varphi_0))\vec{w} +R(\varphi)\vec{w}_0, 
 \sig_0 +\frac 12(1+R(\varphi_0))\vec{w}\times R(\varphi)\vec{w}_0 +\frac 1 2  \sin\varphi_0\vec{w}^2\bigr).
 \end{align}
 We can obtain  adjoint orbits by keeping only linear terms in $\varphi_0,\vec{w}_0, \sig_0$,
 and using $R(\varphi_0)\approx 1 +\epsilon \varphi_0$ for small $\varphi_0$.
 Writing
 \bea
 \xi=\alpha J + \vec{a}\cdot\vec{J} + \tau S,
 \eea 
 and with the convention
 \bea
 \label{vwswitch}
 \vec{v}=-\epsilon\vec{w} \quad (\text{so that} \quad \vec{v}\cdot  \vec{K} = \vec{w}\cdot \vec{ J})
 \eea
 one finds 
\bea
\label{adorbits}
\Ad_v (\xi) = \alpha J +  \vec{a}' \cdot\vec{J} + \tau' S,
\eea
with
\begin{align}
\label{adtrans}
\vec{a}'&=   R(\varphi) \vec{a} +\alpha  \vec{v} \nonumber \\
\tau'& = \tau + \vec{v}\cdot R(\varphi)\vec{a} +\frac 1 2 \alpha  \vec{v}^2.
\end{align}

\subsection{The centrally extended Galilei group }
\label{sec:centr-extend-galil}

  For the calculations in the rest of the paper, we need an explicit definition and description of the centrally extended Galilei group, whose Lie algebra is $\inawi$.  We are going to base our formulation on the observation that the Lie algebra $\inawi$ is a  'complexification' of  the Lie algebra $\nawi$ in the sense defined at the end of Sect.~\ref{gallimit}. Thus we work with 
  formal parameter $\theta$ which satisfies $\theta^2=0$, and use the notation \eqref{nawipara}
  with  entries of the form $a+\theta b$, with $a$ and $b$ real. We denote the two-fold central extension of the Galilei group by $\INW$, and write elements in the form
  \bea
  \label{inawipara}
( \varphi +\theta \alpha, \vec{w} +\theta\vec{a}, \sig +\theta\tau),\qquad \g{a},\tau \in \RR,
 \vec{a} \in \RR^2,
   \eea
   with ranges for $\varphi,\vec{w},\sig$ as defined in \eqref{nawipara}. The element \eqref{inawipara} can be identified with 
    \bea
   \exp(\vec{w}\cdot  \vec{J} + \vec{a}\cdot \vec{P})\exp(\sig S+ \tau M )\exp(\varphi J+ \g{a} H).
  \eea
 This 'complexified' notation turns out be the most efficient for calculations. We  
use
\bea
R(\varphi +\theta \alpha )  = R(\varphi) (1+\epsilon \theta \alpha).
\eea
as well as \eqref{vwswitch} to compute  the multiplication rule from the multiplication rule for  $\NW$: 
\begin{align}
&(\varphi_1 +\theta \alpha_1, \vec{w}_1 +\theta \vec{a}_1, \sig_1 +\theta\tau_1)
(\varphi_2 +\theta \alpha_2, \vec{w}_2 +\theta \vec{a}_2, \sig_2 +\theta\tau_2) \nonumber \\
&=(\varphi_1+\varphi_2 +\theta (\alpha_1+\alpha_2),  \vec{w}_1+R(\varphi_1)\vec{w_2} 
 +\theta( \vec{a}_1 +R(\varphi_1)\vec{a}_2 - \alpha_1\vec{v}_2) , \sig
  +\theta\tau),
\end{align}
where $\sig$ is given as in \eqref{sigmult}, and 
\begin{align}
\tau &= \tau_1+\tau_2  +\frac 1 2 (\vec{a}_1\times R(\varphi_1)\vec{w}_2 +\vec{w}_1\times R(\varphi_1)\vec{a}_2- \alpha_1 \vec{w}_1\cdot  R(\varphi_1)\vec{w}_2) \nonumber \\
&= \tau_1+\tau_2  +\frac 1 2 (-\vec{a}_1\cdot R(\varphi_1)\vec{v}_2 +\vec{v}_1\cdot R(\varphi_1)\vec{a}_2- \alpha_1 \vec{v}_1\cdot  R(\varphi_1)\vec{v}_2).
\end{align}
In studying the representation theory of $\INW$ we will  exploit the fact that $\INW$ is a semi-direct product
\bea
\label{inwsemidirect}
\INW =\NW \ltimes \RR^4.
\eea
  To make this manifest we write elements in factorised form 
   \bea
   \label{facor}
g=x v, \quad \text{with} \quad  x=   (\theta \alpha, \theta \vec{a}, \theta \tilde \tau), \quad v=
(\varphi,\vec{w},\sig),
   \eea
where we used the notation \eqref{inawipara} for both elements.    Note that  purely inhomogeneous  elements  $x$ 
of the group have 'purely imaginary' parameters  $\theta \alpha$ etc..
The factorised form  can be related to  the complex parametrisation \eqref{inawipara} by multiplying out
  \bea
    (\theta \alpha, \theta \vec{a}, \theta \tilde \tau)(\varphi,\vec{w},\sig) =
   (\varphi +\theta \alpha, \vec{w} +\theta (\vec{a} - \alpha \vec{v}), \sig +\theta\tau),
  \eea
   where 
   \bea
   \label{taushift}
  \tau = \tilde \tau + \frac 1 2  \vec{a}\times \vec{w}=\tilde \tau - \frac 1 2  \vec{a}\cdot \vec{v}.
 \eea 
Using \eqref{adtrans} one obtains the following  multiplication law in the semi-direct product formulation 
\begin{align}
&(\theta  \alpha_1, \theta \vec{a}_1, \theta \tilde \tau_1)(\varphi_1,\vec{w}_1,\sig_1)
(\theta \alpha_2, \theta \vec{a}_2, \theta \tilde \tau_2)(\varphi_2,\vec{w}_2,\sig_2) \nonumber \\
 &= (\theta (\alpha_1 + \alpha_2) , \theta (\vec{a}_1+ R(\varphi_1)\vec{a}_2+  
 \alpha_2\vec{v_1}), \theta \tilde \tau)(\varphi_1 +\varphi_2 ,\vec{w}_1 +R(\varphi_1) \vec{w}_2,\sig),
\end{align}
 where $\sig$ is again given by \eqref{sigmult} and 
\bea
\tilde \tau &= \tilde \tau_1 + \tilde \tau_2 + \vec{v}_1\cdot R(\varphi_1)\vec{a}_2 +\frac 1 2 \alpha_2 \vec{v}_1^2.
\eea
Neither the 'complexified' nor the semi-direct product description of  $\INW$ coincide with the 
parametrisation of the 
two-fold central extension of  Galilei group most commonly used in the literature. To make contact with, for example, the parametrisation  in \cite{Mund} we
factorise $\INW$  elements in the form
\bea \label{gal_group}
(\theta \alpha,\vec{0},0)(\varphi ,\vec{w}+\theta \vec{a},\sig
+\theta \tau)  = (\varphi +\g{8} \g{a}, \vec{w} +\theta(\vec{a}-\alpha \vec{v}), \sig +\theta\tau).
 \eea
This almost agrees with the semi-direct product parametrisation
\eqref{facor},  except for the shift \eqref{taushift}. 
The group multiplication rule  in the parametrisation \eqref{gal_group}
% follows from 
%\bea
%(0,\vec{w} +\theta\vec{a},0)(\theta \alpha,\vec{0},0)=(\theta \alpha,\vec{0},0)(0,\vec{w} +\theta(\vec{a} +\alpha \vec{v}),0)
%\eea and 
is 
\begin{align}
&(\theta \alpha_1,\vec{0},0)(\varphi_1,\vec{w}_1+\theta \vec{a}_1,\sig_1+\theta \tau_1) 
(\theta \alpha_2,\vec{0},0)(\varphi_2,\vec{w}_2+\theta \vec{a}_2,\sig_2 +\theta \tau_2) 
\nonumber \\
&=\bigl(\theta (\alpha_1+\alpha_2),\vec{0},0\bigr)\bigl((\varphi_1+\varphi_2),\vec{w}_1 +
R(\varphi_1)\vec{w}_2+\theta (\vec{a}_1+R(\varphi_1)\vec{a}_2+ \alpha_2\vec{v}_1),\sig +\theta \tau\bigr), 
\end{align}
where $\sig$ is again \eqref{sigmult} and 
\begin{align}
 \tau &=  \tau_1 +  \tau_2 +\frac 1 2 (\vec{a}_1\times R(\varphi_1)\vec{w}_2 + \vec{w}_2 \times R(\varphi_1)\vec{a}_2 + \alpha_2\vec{v}_1\times  \vec{w}_2)
 \nonumber \\
  &=  \tau_1 +  \tau_2 -\frac 1 2 (\vec{a}_1\cdot R(\varphi_1)\vec{v}_2 - \vec{v}_1 \cdot R(\varphi_1)\vec{a}_2 + \alpha_2\vec{v}_1\cdot  R(\varphi_1) \vec{v}_2).
\end{align}
This matches the conventions used for the centrally extended Galilei group in \cite{Mund}  up to a sign $\tau \rightarrow -\tau$.

\subsection{Coadjoint orbits as adjoint orbits and the phase spaces of Galilean particles}

 \label{adorbsec}
The  coadjoint orbits  of the centrally extended Galilei group $\INW$
were studied in detail in \cite{BGdO,NdOT}.  For our purposes it is
essential to identify the coadjoint orbits with 
 adjoint orbits, using the inner product \eqref{inprod2}.  In that way, we will be able to 
 make contact with the group conjugacy classes which play an important role later in this paper.
 Thus we identify the generic element in the dual of the  Lie algebra
 \bea
 \label{genliedual}
 Q^*= m M^* + E H^* +  \anj J^* + s S^*+ \vec{p}\cdot \vec{P}^* +
 \vec{j}\cdot\vec{J}^*
  \eea
  with the Lie algebra element
  \bea
  \label{genlie}
  Q= m J + E S +  \anj M + sH -\vec{p}\cdot \vec{J} -
 \vec{j}\cdot\vec{P}.
  \eea
 In order to describe adjoint orbits explicitly, we pick a reference point $Q_0$ in $\inawi$, and write a generic point in the orbit as  
  \bea
  \label{genorb}
  Q=\Ad_g(Q_0),
  \eea
  where  we parametrise $g\in\INW$ 
 as is customary  in the literature referring to the Galilei group \eqref{gal_group}:
\bea
\label{inwpara}
g = (\theta \alpha, \vec{0}, 0)(\varphi, \vec{w} + \g{8} \vec{a}, \sig
+ \g{8} \g{t}) =  (\varphi +\g{8} \g{a}, \vec{w} +\theta(\vec{a}-\alpha \vec{v}), \sig +\theta\tau).
\eea
 
 Before we carry out the computation we should clarify the physical interpretation of the various parameters we have introduced.  A coadjoint orbit of the Galilei group (or its central extension) can be interpreted as the  phase space of a Galilean particle. The coordinates on the phase space parametrise the states of motion of such a particle. A full justification of the interpretation  requires  Poisson structure of the phase space, which is discussed in the literature \cite{BGdO,NdOT} and which we review  in  Sect.~\ref{poisssect}. However,  we give the  physical picture here  so that we can interpret the formulae we derive as we go along.  

The vectors $\vec{p}$ and $\vec{j}$ are the particle's momentum and impact vector (the quantity which is conserved because of invariance under Galilei boosts).    The parameters $m$ and $s$ are the mass and spin of the particle \cite{JN}, 
$E$ is the energy and $\anj$  the angular momentum.  Note that $s$ is related to $\anj$
in the rest frame of the particle  in exactly the same way in which $m$ is related to $E$ in that frame \cite{JN2}. 
The  interpretation of the  parameters in the group element \eqref{inwpara} follows from our discussion of the Galilei group   in Sect.~\ref{galfoundations}: the vector
$(\alpha,\vec{a})\in \RR^3$  parametrises spacetime translations, the
angle $\varphi$ parametrises spatial rotations, and $\vec{w}$ is
related to the boost velocity via \eqref{vwswitch}.  The parameters $
\tau $ and $ \sig $ do not appear to have a clear geometrical significance.

Our strategy for computing these orbits is to start with adjoint orbits  of the homogeneous group 
$\NW$, and to obtain orbits in the full inhomogeneous group by the complexification trick described in  Sect.~\ref{sec:centr-extend-galil}. Thus we start with an element
  \bea
 Q_0 = m J - \vec{p}_0\cdot \vec{J} +E_0 S
 \eea
 in the Lie algebra of the homogeneous group $\NW$. 
 We already know the  effect of conjugating with an element $v= (\varphi, \vec{w}, \sig)\in \NW$
from our earlier calculation \eqref{adorbits}.  In the current notation the result reads  
 \bea
 \label{adnw}
 \Ad_v (Q_0)= mJ   - \vec{p}\cdot \vec{J} + E S,
 \eea
 with 
 \begin{align}
 \label{nworb1}
 \vec{p}&=  R(\varphi)\vec{p}_0 -m\vec{v} \nonumber \\
 E &= E_0 -  \vec{v}\cdot R(\varphi)\vec{p}_0 +\frac 1 2 m \vec{v}^2.
 \end{align}
  Physically, the formulae \eqref{nworb1} give the behaviour or energy
  and momentum under a boost, translation and rotation of the
  reference frame.  We will need to know the different kinds of orbits  and their associated centralisers  when we study the representation theory of
  $\INW$.

   Trivial orbits arise when we start with a reference element for which both mass and spin vanish:
  \bea
  \label{qe}
 Q_0(E_0) = E_0 S.
 \eea
The orbits consists of  point $ O_{E_0}=\{E_0S\}$; the associated centraliser group is the entire group  i.e. $N_{E_0}=\NW$.
When the mass parameter $m\neq 0$ i.e.
 \bea
 \label{qem}
 Q_0(E_0,m) =E_0S+ m J , \qquad m\neq 0
  \eea
the orbit  is 
\bea
\label{paraorbit}
O_{E_0,m}=\{mJ -\vec{p}\cdot \vec{J} + ES|E=E_0+ \frac {1} {2m}\vec{p}^2\}, 
\eea
which is a paraboloid opening towards the positive or negative $E$-axis, depending on the 
sign of $m$.
The centraliser group is 
\bea
\label{paracent}
N_{E_0,m}= \{v=(\varphi,\vec{0},\sig)\in \NW | \varphi \in [0,2\pi), \sig \in \RR\}.
\eea
Finally there are non-trivial orbits when the mass is zero, but the momentum non-zero.  The 
reference element
\bea
\label{qp}
 Q_0(\vec{p}_0) =-\vec{p}_0\cdot\vec{J}
 \eea
leads to the orbit  
\bea
\label{cylorbit}
O_{\vec{p_0}}=\{ES -\vec{p}\cdot \vec{J}||\vec{p}| =|\vec{p}_0|, E\in \RR\},
\eea
which  is a cylinder. The centraliser group is 
\bea
\label{cylcent}
N_{\vec{p_0}} = \{v= (0,\lambda \vec{p}_0, \sig)|\lambda,\sig \in \RR\}.
\eea

 Moving on to the inhomogeneous group, we  write a generic but fixed element in  the Lie algebra $ \inawi$ in 'complexified' language:
 \begin{equation}
\begin{split}
 Q_0 = (m + \g{8} s) J - \bigl(\vec{p}_0 + \g{8}\vec{j}_0\bigr) \vec{J} + (E_0 + \g{8}\anj_0)S.
 \end{split}
\end{equation}
In order to compute the effect of conjugating this element with an element in $\INW$ parametrised as in \eqref{inwpara} we simply 'complexify' the result \eqref{nworb1} i.e. we replace
$\varphi$ by $\varphi +\alpha\g{8}$,  $\vec{p}_0$ by $\vec{p}_0+\g{8}\vec{j}_0$ etc.
Parametrising the result as 
 \bea
 \label{complexorbit}
\Ad_g( Q_0) = (m' + \g{8} s') J  -\bigl(\vec{p} + \g{8}\vec{j}\bigr) \vec{J} + (E + \g{8}\anj)S
 \eea
 we see straight away   that $m=m'$ and $s=s'$. The formulae  
  $\vec{p}$ and $E$ are  simply the terms of order zero in $\g{8}$, and therefore given by the expressions in  \eqref{nworb1}. In order to find $\anj$ and $\vec{j}$ we need to compute the terms which are linear in $\theta$. Thus $\anj$  is the coefficient of $\theta$ in 
 \begin{equation}
   \g{8}\anj_0 +\g{e}\bigl(\vec{w} + \g{8} (\vec{a} - \g{a}\vec{v})\bigr)
\cdot R(\g{ff})(1 + \g{e}\g{8} \g{a})(\vec{p}_0 + \g{8}\vec{j}_0)  +  \dfrac{1}{2} (m + \g{8} s)\bigl(\g{e}\vec{w} +  \g{e}\g{8} (\vec{a} -
\g{a}\vec{v}) \bigr)^2.
 \end{equation}
 Multiplying out  and simplifying we find
  \bea
  \label{orball:2}
   \anj & =
 \anj_0 - \vec{v}
\cdot R(\g{ff})\vec{j}_0+ \g{e} \vec{a} \cdot
R(\g{ff})\vec{p}_0 + \dfrac{1}{2} s \vec{v}^2 + m \vec{a}\times \vec{v}.
\eea
In order to compute $\vec{j}$, we need the coefficient of $\theta$ in 
\begin{equation}
R(\g{ff})(1 + \g{e}\g{8}\g{a}) \cdot ( \vec{p}_0 +
\g{8} \vec{j}_0) +  (m + \g{8} s)\g{e}\bigl(\vec{w} +\g{8} (\vec{a} -
\g{a}\vec{v}) \bigr), 
\end{equation}
which is 
\bea
\label{orball:3}
\vec{j} =  R(\g{ff})\vec{j}_0 +\g{a}  \g{e} R(\g{ff})\vec{p}_0+
 m \g{e}\bigl( \vec{a}  -\g{a} \vec{v}\bigr)  -s \vec{v}.
\eea

Having obtained explicit expressions for the adjoint
orbits of $ \inawi $  it is straightforward to classify the different types of orbits which arise.
This classification essentially follows from the classification of conjugacy classes in the  homogeneous group $\NW$ in   \cite{FoFS} by linearising and using the  
'complexification'  trick to extend to the inhomogeneous situation. These orbits are already discussed in the literature \cite{BGdO,NdOT}, albeit in a different notation. We therefore  simply list the orbit types in our notation in Table~\ref{inawiorbits}, giving in each  case an element 
from which the orbit is obtained by 
conjugation.  The non-trivial orbits are  two- or four dimensional. The two-dimensional orbits are essentially the adjoint orbits of $\NW$, which return here in the inhomogenous part of $\INW$.  For us, the two four-dimensional orbits are most relevant: they correspond to, respectively,  massive and massless particles.

%Equations  \eqref{nworb1},~\eqref{orball:2} and \eqref{orball:3} give the
%coadjoint orbits in $ \inawi$.\footnote{ Note that this agrees with the results
%in \cite{NdOT} but the notation is different; we use  $
%R(\g{ff})\vec{a} $ in place of $  \vec{a}^{~\g{ff}} $ that is used
%there. Also there is a relative rotation by $ \g{p} $
%between the two notations so our tuple $ (\vec{j}, \vec{v}, \vec{p}, \vec{a})
%$is related to their $ (\vec{\mathrm{k}}, \vec{\mathrm{v}}, \vec{\mathrm{p}}, \vec{\mathrm{a}}) $ in a manner similar to \eqref{vwswitch}
%\begin{equation*}
%  \label{eq:2}
%  (\vec{\mathrm{k}}, \vec{\mathrm{v}}, \vec{\mathrm{p}},
%  \vec{\mathrm{a}}) =  (\vec{j}, \g{e}\vec{v}, -\g{e}\vec{p}, \g{e}\vec{a}).
%\end{equation*}}

\begin{table}[h! b! p!]
\baselineskip 30 pt
\begin{center}
\begin{changemargin}{-.5cm}{0cm}
\centering
\begin{tabular}{|c| p{7.0cm}| p{3.4cm} p{0.01cm}|}
\hline
%  &    &   \\
 \centering \textbf{Reference element $Q_0$}
 
 & \centering \textbf{Generic element $Q$ on  orbit}

 &\centering \textbf{Orbit geometry} & \\
%constant & & \\
 % &    &   \\
\hline
%  &    &   \\
$ E_0 S + j_0 M $  & \centering $E_0 S + j_0 M$ & \centering \text{Point} &\\
% &    &   \\
\hline
 % &    &   \\
$E_0 S+  j_0 M + s H $ & \centering
 $ E_0 S+ j M + s H - \vec{j}\cdot\vec{P}$ \newline with $ \vec{j}
 \in \mathbb{R}^2, j = j_0 +
 \tfrac{1}{2s}\vec{j}^2$ &\centering
Paraboloid$\simeq \RR^2$ &\\
\hline

 % &    &   \\
$  E_0S +j_0 M - \vec{j}_0\!\cdot\!\vec{P}$ & \centering
 $ E_0 S+ j M - \vec{j} \cdot\vec{P}$ \newline with 
 $ j\in \RR, |\vec{j}| = |\vec{j}_0| $          & \centering
 Cylinder $\simeq S^1\times \RR$&\\

\hline
 % &    &   \\
$E_0S+ m J + j_0 M + s H$ & \centering
 $ES + m J -\vec{p}\cdot \vec{J} + j M -
 \vec{j}\cdot\vec{P}+ s H $
 \newline with $\vec{p},\vec{j}\in\RR^2$,  $E=E_0+ \frac {1} {2m}\vec{p}^2, j =
 j_0 + \tfrac{1}{m}\vec{p}\cdot \vec{j} - \tfrac{s}{2 m^2} \vec{p}^2$ & \centering
 Tangent bundle of paraboloid$\simeq \RR^4$ &\\
 % &    &   \\
\hline
 % &    &   \\
$ E_0 S -\vec{p}_0\!\cdot\!\vec{J} + j_0 M -
\vec{j}_0\!\cdot\!\vec{P} +sH$ & \centering
 $ ES -\vec{p}\cdot \vec{J} + j M - \vec{j}\cdot\vec{P}+sH $
 \newline with $ E, j \in \RR,  |\vec{p}| =|\vec{p}_0|, \newline
 \vec{j}\cdot \vec{p}-Es= \vec{j}_0\cdot \vec{p}_0-E_0s_0$ & \centering
 Tangent bundle of cylinder $\simeq S^1\times \RR^3$&\\
 % &    &   \\
\hline

\end{tabular}
\end{changemargin}
\end{center}
\caption{ \label{inawiorbits} Adjoint orbits of the centrally extended Galilei group $\INW$}
\end{table}

\subsection{Poisson structure of the orbits}
\label{poisssect}
The canonical one-form associated to a coadjoint $\INW$-orbit labelled by one of the 
reference elements $
Q_0 $ in Table~\ref{inawiorbits} is of the general form \cite{Kirillov}
\bea 
\label{canform}
\boldsymbol{\g{8}}_{Q_0}=\lie{Q_0}{g^{-1}\dd
  g}.
\eea
To compute  this, we first  note the  left-invariant Maurer-Cartan one-form on $\NW$ in our parametrisation (see also \cite{Nappi:1993ie}, where this was first computed, using different coordinates). With the notation \eqref{vnotation}, it is
\bea
v^{-1}\dd v  = R(-\g{ff}) \dd\vec{w} \cdot \vec{J} + \dd\g{ff} J + (\dd
\sig + \dfrac{1}{2} \dd\vec{w} \times \vec{w}) S.
\eea
For later use we also record the expression for the right-invariant
Maurer-Cartan form on $\NW$
\bea
\label{rmc}
\dd v v^{-1} & = (\dd\vec{w} -\dd\varphi \epsilon \vec{w} )\cdot \vec{J} + \dd \g{ff} J + (\dd
\sig  +\frac 1 2 \vec{w}^2 \dd\varphi- \dfrac{1}{2} \dd\vec{w} \times \vec{w}) S.
\eea
Next, we again use the 'complexification' trick to  obtain  the left-invariant Maurer-Cartan form on the centrally extended Galilei group $ \INW $, which appears in \eqref{canform}.  In terms  of the  parametrisation  \eqref{gal_group}
\[
g= (\varphi +\g{8} \g{a}, \vec{w} +\theta(\vec{a}-\alpha \vec{v}), \sig +\theta\tau)
\]
we find
%\begin{equation}
%\begin{split}
%g^{-1}\dd g & = R(-\g{ff})(1 - \g{e}\g{8}\g{a})\dd \bigl(\vec{w} +
%\g{8}(\vec{a} - \g{a} \vec{v}) \bigr) \cdot \vec{J} + \dd (\g{ff} +
%\g{8}\g{a}) J \\
%&+ \bigl[\dd (\sig + \g{8} \g{t}) + \dfrac{1}{2} \dd\bigl(\vec{w} +
%\g{8}(\vec{a} - \g{a} \vec{v}) \bigr) \times \bigl(\vec{w} +
%\g{8}(\vec{a} - \g{a} \vec{v}) \bigr)\bigr] S,
%\end{split}
%\end{equation}
%which simplifies to
\begin{equation}\label{mcinw}
\begin{split}
g^{-1}\dd g = &R(-\g{ff}) \dd\vec{w} \cdot \vec{J} + \dd \g{ff} J + (\dd
\sig + \dfrac{1}{2} \dd\vec{v} \times \vec{v}) S ~ + R(-\g{ff}) \bigl(\dd\vec{a} - \vec{v}\dd\g{a}\bigr)\cdot \vec{P}\\
& +
\dd\g{a} H + \bigl(\dd \g{t} + \frac{1}{2} \vec{v}^2 \dd\g{a}
+ \dfrac{1}{2} \dd(\vec{v} \cdot \vec{a}) - \vec{v}\cdot \dd\vec{a}\bigr) M.
\end{split}
\end{equation}
It is explained in \cite{MS2}, that for semi-direct product groups of the type $H\ltimes \mathfrak{h}^*$, there is an alternative  formula for  the symplectic structure on coadjoint orbits, which will be useful when we study Poisson-Lie group analogues of coadjoint orbits.  In this alternative expression  one pairs the inhomogeneous part of  a generic point on the orbit with the right-invariant Maurer-Cartan form \eqref{rmc} on the homogeneous part of the group. In our case, this gives
\bea
\label{adpotential}
\tilde{\boldsymbol{\g{8}}}_{Q_0}=\langle  sH-\vec{j}\cdot\vec{P}+jM) , \dd v v^{-1}\rangle,
\eea
where we assume the expressions  for $\vec{j}$ and $\anj$ given in Table~\ref{inawiorbits} for the generic orbit element $Q$ on the orbit with reference element $Q_0$.

We briefly illustrate the formula for the symplectic potential by evaluating it for the two four-dimensional orbits in Table~\ref{inawiorbits}.
Inserting the  element $Q_0$ representing  a massive particle with spin
we obtain the symplectic potential for the corresponding orbit:
\begin{equation}\label{can_one_form}
\begin{split}
\boldsymbol{\g{8}}_{E_0,j_0,m,s}& = \lie{E_0S+mJ+j_0M+sH}{g^{-1}\dd g} = \\
& = \anj_0 \dd \g{ff}  +  s (\dd
\sig + \dfrac{1}{2} \dd\vec{v} \times \vec{v}) + E_0\dd\g{a} + m \bigl(\dd \g{t} + \frac{1}{2} \vec{v}^2 \dd\g{a}
+ \dfrac{1}{2} \dd(\vec{v} \cdot \vec{a}) - \vec{v}\cdot \dd\vec{a}\bigr),
\end{split}
\end{equation}
which simplifies to 
\bea
\label{massivepot}
\boldsymbol{\g{8}}_{E_0,j_0,m,s} = \dfrac{s}{2} \dd\vec{v} \times \vec{v} + m\vec{v}\cdot\dd(\alpha \vec{v}-\vec{a}) \quad \text{+ exact term}.
\eea
The associated symplectic form   only depends on  $m$ and $s$, and is given by
\begin{equation}\label{symplectic_form}
\g{w}_{ms} = \dd\boldsymbol{\g{8}}_{E_0,j_0,m,s} = \dfrac{s}{2} \dd(\g{e}\vec{v}) \wedge \dd\vec{v}
+ m \vec{v} \cdot \dd\vec{v} \wedge \dd\g{a} - m \dd\vec{v} \wedge \dd\vec{a},
\end{equation}
where we use the shorthand $d\vec{a}\wedge \dd\vec{b}=da_i \wedge \dd b_i$.
In terms of  the momentum vector $\vec{p}=-m\vec{v}$ we recover  the fomula given in 
\cite{JN}:
\begin{equation}
  \label{eq:4omega}
  \g{w}_{m,s} = \dfrac{s}{2 m ^2}\dd(\g{e}\vec{p}) \wedge \dd\vec{p}
- \dfrac{1}{m} \vec{p} \cdot \dd\vec{p} \wedge \dd\g{a} + \dd\vec{p} \wedge \dd\vec{a}.
\end{equation}
%Thus a possible set of  canonical coordinates  is
%\begin{align}
%\vec{Q} &:=  \vec{a}- \dfrac{1}{m}\g{a} \vec{p}+\dfrac{s}{2 m ^2} \g{e}\vec{p} 
%\\
%\vec{P} &:= \vec{p}.
%\end{align}
%\rem{ check sign conventions in above: should we use $\vec{p}=-m\vec{v}$ instead?}
Computing  the symplectic potential via the alternative expression \eqref{adpotential} we find 
\bea
\tilde{ \boldsymbol{ \g{8} } }_{E_0,j_0,m,s}=
\anj_0 \dd \g{ff}  +  s (\dd
\sig + \dfrac{1}{2} \dd\vec{v} \times \vec{v})  + m \bigl( \frac{1}{2} \vec{v}^2 \dd\g{a}
+  \vec{a}\cdot \dd\vec{v}\bigr), 
\eea
which clearly only differs from $\boldsymbol{ \g{8} }_{ms}$ by exact terms, and leads to the same symplectic form.

Proceeding similarly for the four-dimensional orbit corresponding to massless particles, we  compute
\bea
\boldsymbol{\g{8}}_{E_0,j_0,\vec{\pi}_0,\vec{j}_0,s} = \lie{E_0S-\vec{\pi_0}\cdot\vec{J}-\vec{j}_0\cdot\vec{P}+j_0M+sH}{g^{-1}\dd g}
\eea
and find
that the  symplectic potential  is
\bea
\boldsymbol{\g{8}}_{E_0,j_0,\vec{\pi}_0,\vec{j}_0,s} =
(\vec{j} +\frac s 2\vec{v})\cdot \dd(\epsilon\vec{v})+\dd \vec{p}\cdot (\alpha \vec{v}-\vec{a}) 
\quad \text{+ exact term},
\eea
where it is understood that $\vec{p}$ and $\vec{j}$ are given by the expressions \eqref{nworb1} and 
\eqref{orball:3}.  Thus we  obtain  the symplectic potential 
\bea
\omega_{\vec{p}_0, \vec{j}_0,s}=\dd \vec{j} \wedge \dd(\epsilon\vec{v}) +\frac s 2 \dd \vec{v}\wedge\dd(\epsilon \vec{v}) - \alpha \dd\vec{p}\wedge \dd\vec{v} - \vec{v}\cdot  \dd\vec{p}\wedge \dd \alpha
+ \dd\vec{p}\wedge\dd\vec{a}.
\eea

\section{Galilean gravity in three dimensions  as gauge theory of the centrally extended Galilei group}

\subsection{The Chern-Simons formulation of Galilean gravity}

Our approach to formulating classical and, ultimately, quantum theory of Galilean gravity is based on the Chern-Simons formulation of three-dimensional gravity discovered  in \cite{AT} and elaborated in \cite{Witten} for general relativity in three dimensions.  The prescription is, very simply, to pick a differentiable three-manifold $\manif$ describing the universe and to write down the Chern-Simons action for the isometry group of the model spacetime for a given signature and value of the cosmological constant, using an appropriate inner product on the Lie algebra.  Such an action is, by construction, invariant under orientation-preserving diffeomorphisms of   $\manif$. The prescription makes sense for any  differentiable  three-manifold $\manif$, but for the Hamiltonian quantisation approach  we need to  assume $\manif$  to be of topology $\RR\times \Sigma $, where $\Sigma$ is a two-dimensional manifold, possibly with punctures and a boundary.

For our model of Galilean gravity  we use the extended  Galilei group $\INW$ as a gauge group, equipped with invariant inner product \eqref{inprod1}.
The  gauge field of the Chern-Simons action is then  locally given by a one-form on $\manif$ with values in the Lie algebra  of the centrally extended Galilei group:
 \bea
 A = \omega  J + \omega_i K_i  + \eta S + e H + e_i P_i + f M,
 \eea
 where $\omega, \omega_1, \omega_2, \eta, e,  e_1, e_2$  and $f$ are ordinary one-forms on $\manif$.  For later use we split this into a homogeneous part
 \bea
 a= \omega  J + \omega_i K_i  + \eta S
 \eea
  and a normal part
  \bea
b=   e H + e_i P_i + f M.
  \eea
 Then one checks that 
 \bea
\frac 1 2  [A,A] = \epsilon_{ij} \omega_i\wedge \omega  K_j + \omega_1\wedge\omega_2 S
+( \epsilon_{ij}e_i\wedge \omega + \omega_j\wedge e)  P_j +\omega_i\wedge  e_i M.
 \eea
 The curvature  
 \bea
 \label{curvsplit}
 F=\dd A +\frac 1 2 [A,A]  = R +T
 \eea
 is a sum of the homogeneous terms, taking values in $\inawi$,
 \bea
R= \dd\omega \; J+ (\dd\eta + \omega_1\wedge \omega_2) S +(\dd\omega_i +\epsilon_{ij}\omega\wedge \omega_j) K_i
\eea
and the  torsion part
\bea
T = \dd e\; H  + (\dd f +\omega_i\wedge e_i) M +(\dd e_j +\omega_j\wedge e + \epsilon_{ij}e_i\wedge \omega)P_j.
\eea
 
 Thus the Chern-Simons action for Galilean  gravity in 2+1 dimensions takes the form
 \begin{align}
 I_{CS}[A]&=\frac{1}{16\pi G} \int_\manif \langle A\wedge \dd A\rangle +\frac 1 3 \langle A\wedge[A,A]\rangle 
 \nonumber \\
 &=\frac{1}{16\pi G}\left[ \int_\manif \omega \wedge \dd f + f\wedge
   \dd\omega + e\wedge \dd\eta + \eta\wedge \dd e +\epsilon_{ij}
   (\omega_i\wedge \dd e_j + e _j\wedge \dd \omega_i)\right. \nonumber \\
   &+ 2\left. \int_\manif\omega \wedge \omega_i\wedge e_i + e\wedge \omega_1\wedge \omega_2\right],
 \end{align}
 where $G$ is Newton's constant in 2+1 dimensions, with physical dimension of  inverse mass.  With  both the connection $A$ and the curvature being dimensionless, the physical dimension length$^2$/time of the pairing $\langle, \rangle$ combines with the dimension mass of $1/G$ to produce the required dimension of an action. 
 
 As expected for gauge groups of the semi-direct product form, with a pairing between the homogenous and the normal part, this action is equivalent to  an action of the  BF form,
 \begin{align}
 I_{BF} [A]&=\frac{1}{8\pi G}\int_\manif \langle R\wedge b\rangle  \nonumber \\
 &= \frac{1}{8\pi G}\int_\manif \dd\omega \wedge f + \dd\eta \wedge e + \epsilon_{ij} \dd\omega_i\wedge e_j
 + e\wedge \omega_1\wedge \omega_2 + \omega \wedge \omega_i\wedge e_i,
 \end{align}
 with $I_{CS}$ and $I_{BF}$ differing by an integral over an exact form i.e. a  boundary term.  This BF-action is closely related to the action of lineal gravity \cite{CJ}: the two actions are based on the same curvature $R$ and the same inner product \eqref{NWinprod}. The difference is that the one-form $b$ is  a Lagrange multiplier {\em function} $\eta$ in  the gauge formulation of  lineal gravity.

 In the absence of a  boundary and  punctures on $\Sigma$  
  the stationary points of the  Chern-Simons action   are  flat connections i.e. connections for which 
  \bea
  \label{Galgraveq}
  F=0 \Leftrightarrow R=T=0.
  \eea
  These are the vacuum field equations of classical Galilean gravity.
     The interpretation of  solutions,  i.e.  flat $\INW$-connections,  in terms of Galilean  spacetimes proceeds along the same lines as the interpretation of flat Poincar\'e group connections in terms of Lorentzian spacetimes, see e.g. \cite{DJtH,SousaGerbert,MS4}. The flatness condition means that the gauge field can be written in local patches as $A=g^{-1} \dd g$ for a $\INW$-valued function $g$ in a given patch. Parametrising $g$ in the semi-direct product form \eqref{facor} $g=(\theta T,\theta \vec{X}, \theta\tau)(\phi,\epsilon\vec{V},\sig)$ the $\RR^3$-valued function $(T,\vec{X})$ provides an embedding of the patch into our model Galilean space
  $A^3$ (which we can identify with $\RR^3$ after a choice of origin and frame). 
 In this way, a flat $\INW$ connection translates into an identifications of open sets in $\manif$ with open sets in $A^3$;  overlapping open sets are  connected by  Galilean transformation.
 Note that   coordinates $\tau$ and $\sig$ referring to the central generators play no role in the geometrical interpretation.
  At this level, there is remarkably little difference between relativistic and Galilean gravity. 
 In particular, even though the model space $A^3$ has an absolute time (up to a global  shift), solutions  of the Galilean vacuum field equations \eqref{Galgraveq} need not have an absolute time. In particular there can be time shifts when following the identifications  along  non-contractible paths.  
 
 A  further   important consequence of the field equations \eqref{Galgraveq}
  is that a given  infinitesimal diffeomorphism can be written as  infinitesimal gauge transformation, as in relativistic 3d gravity \cite{Witten}.  This follows from Cartan's identity applied to the $\INW$-connection $A$:
 \bea
  {\mathcal L}_\xi A = \dd \iota_\xi A + \iota_\xi \dd A = D_A(\iota_\xi A),
 \eea 
  where we used the flatness  condition $\dd A=-A\wedge A$ to write the infinitesimal diffeomorphism    ${\mathcal L}_\xi$  as  an infinitesimal  gauge transformation with generator
 $\iota_\xi(A)$.  Conversely we find, as in \cite{Witten},  that an infinitesimal local  translation can be written as local diffeomorphisms provided the frame field $eH + e_iP_i$ is invertible. If that condition is met,  local translations  do not lead to (undesirable) gauge invariance in addition to local diffeomorphisms and homogeneous Galilei transformations.   Questions regarding the role of non-invertible frames and  the relation between global diffeomorphisms  and large gauge transformations are difficult, as they are in the relativistic case, see  \cite{Matschull} for a discussion and further references.   However, unlike in the relativistic case, the issue of understanding the difference between the Chern-Simons and metric formulation does not arise
since  we do not have a metric theory of Galilean gravity in three dimensions. 
  
  In this paper we will not consider boundary components of $\Sigma$, although the boundary treatment in \cite{MS5} includes the current situation as  a special case. We will, however, couple particles to the gravitational field. This is done by marking points on the surface $\Sigma$ and decorating them with coadjoint orbits of the extended Galilei group, equipped with the symplectic structures we studied in Sect.~\ref{poisssect}.
In order to couple the bulk action to particles we need to switch notation which makes the 
splitting $\manif=\RR \times \Sigma $  explicit. 
Referring the reader to \cite{MS4,MS5} for details,
we introduce a coordinate $x^0$ along $\RR$  and coordinates $x^1,x^2$ on $\Sigma$, and split the gauge field into 
\bea
A=A_0\dd x^0 + A_\Sigma,
\eea
where $A_\Sigma$ is a one form on $\Sigma$ (which may depend on $x^0$). For simplicity we consider only one particle, which is modelled by a single puncture of $\Sigma$ at $\vec{x}_*$,  decorated by a coadjoint, or equivalently adjoint,  orbit. In the 
parametrisation of adjoint orbits  of the  centrally extended Galilei in \eqref{genorb}, the group element $g\in \INW$  now becomes a function of $x^0$. 
 The combined field and particle action, with $Q$ and $Q_0$ defined as in \eqref{genorb},   takes the following form:
\begin{align}
 I_\tau[A_\Sigma,A_0, g]=
\label{ordaction}
 &\frac{1}{8\pi G} \int_\RR \dd x^0\int_{\Sigma}\langle \partial_0 A_\Sigma \wedge A_\Sigma\rangle
 -\int_\RR \dd x^0 \langle  Q_0 \,,\, g^{-1}\partial_0
g\rangle \\
 +&\int_\RR \dd x^0\int_{\Sigma}\langle A_0\,,\, \frac{1}{8\pi G} F_\Sigma -
 Q \delta^{(2)}(\vec{x}-\vec{x}_*)\dd x^1\wedge \dd x^2 \rangle.
\nonumber
\end{align}
 Varying with respect to the Lagrange multiplier $A_0$
 we obtain the constraint
\begin{align}
\label{vara0equ} F_\Sigma(x)= 8\pi G Q
\delta^{(2)}(\vec{x}-\vec{x}_*)\dd x^1\wedge \dd x^2.
\end{align}
Variation with respect to $A_\Sigma$ gives  the evolution equation
\bea
\label{aeom}
\partial_0 A_\Sigma=\dd_\Sigma A_0+[A_\Sigma,A_0].
\eea
The equation obtained by varying $g$ does not concern us here, but can be found e.g. in \cite{MS5}.

Together with \eqref{vara0equ} this means that the 
curvature $F$ vanishes except at the 'worldline' of the puncture,
where it is given by the constraint \eqref{vara0equ}.
In order to interpret the constraint
  geometrically and physically 
we  use the decomposition \eqref{curvsplit} of the curvature
and parametrisation \eqref{genlie}:
\begin{align}
R_\Sigma^a&=8\pi G (m J + E S  -\vec{p}\cdot \vec{J})
\delta^{(2)}(\vec{x}-\vec{x}_*)\dd x^1\wedge \dd x^2\nonumber \\ T_\Sigma^a&=8\pi G( sM + \anj H -
 \vec{j}\cdot\vec{P})
\delta^{(2)}(\vec{x}-\vec{x}_*)\dd x^1\wedge \dd x^2.
\end{align}
The constraints mean that the holonomy around  the puncture is related to the  Lie algebra element $Q$ via
\bea
\label{inhomoholo}
h=\exp(8\pi GQ),
\eea
where we have implicitly chosen a reference point where the holonomy begins and ends.  We will return to the role of this reference point when we discuss the phase space of the theory further below.
Since the element $Q=gQ_0g^{-1}$ lies in a fixed adjoint orbit, the holonomy has to lie in a fixed conjugacy class  the element $\exp(Q_0)$. As we saw in Sect.~\ref{adorbsec}, the fixed element $Q_0$ combines the physical attributes of a point particle, like mass and spin.  The holonomy  $h$ thus lies in fixed conjugacy  class labelled by those physical attributes. 
As a preparation for discussing the phase space of the Chern-Simons theory  we therefore study the geometry of conjugacy classes in $\INW$ and their parametrisation in terms of physically meaningful parameters.

\subsection{Conjugacy classes of in the centrally extended Galilei group}

\label{conjugacysect}

 We begin with a brief  review  of the conjugacy classes in $\NW$ in our notation. The classification of conjugacy classes of $\NW$  is given in \cite{FoFS}, and follows from  general formula \eqref{ccformula}. We would like to interpret this classification in terms of Galilean gravity.  The key formula here is \eqref{inhomoholo}. Denoting the $\NW$-valued part of the holonomy $h$  by $u$ we obtain 
\bea
u=\exp(8\pi G (m J + E S  -\vec{p}\cdot \vec{J})).
\eea

In order to understand the relation between the adjoint orbits  discussed after \eqref{adnw}  and conjugacy classes in $\NW$,  we rescale the physical parameters mass, energy and momentum which label the  adjoint orbits and define  
\bea
\label{scalepara}
\mu = 8\pi Gm, \quad \grp_0= 8\pi G\vec{p}_0, \quad \gren_0= 8\pi G E_0.
\eea
Then we define group elements corresponding to the  special Lie algebra elements  \eqref{qem}  and \eqref{qp} via
 \bea
 \label{homorefs}
u_0(\gren_0)=(0,\vec{0}, \gren_0), \qquad
u_0(\mu,\gren_0)=(\mu,\vec{0},\gren_0), \qquad u_0(\grp_0)=(0,-\grp_0,0),
\eea
where we assume $\mu\neq 0,2\pi,\grp_0\neq \vec{0}$.
When parametrising the $\NW$-conjugacy classes obtained by conjugating these elements we could  either use parameters $E,\vec{p},m$  in the Lie algebra or a group parametrisation of the form 
\bea
\label{homopara}
u=(\mu,-\grp,\gren). 
\eea
The two are  related via
\bea
\label{grouplie}
(\mu,-\grp,\gren)= \exp(8\pi G(mJ-\vec{p}\cdot \vec{J} + ES)).
\eea
We exhibit the relationship between the two parametrisations for each  type of conjugacy class separately. As we shall see, we can think of $\gren$ and $\grp$ as  energy and momentum parameters on the group, and $\mu$ as an angular mass coordinate, with range $[0,2\pi]$. 
The compactification of the range of the mass from  an infinite range to a finite interval is 
a familiar feature of three-dimensional gravity, in both the Lorentzian and Euclidean regime 
\cite{tHooft,MatWell}.

For the trivial conjugacy classes  consisting of $u_0(\gren_0)$  we have simply $\gren_0=8\pi G E$. 
  The conjugacy class consists  of  a point $ C_{\gren_0}=\{(0,\vec{0},\gren_0)\}$; the associated centraliser group is the entire group  i.e. $N_{\gren_0}=\NW$.
The conjugacy class containing $u_0(\mu,\gren_0)$ can be parametrised as 
\bea
\label{uconj1}
C_{\gren_0,\mu}=\{(\mu,-\grp,\gren) |\gren =\gren_0+\frac 1 4 \cot\left(\frac \mu 2\right)\grp^2 \}.
\eea
Provided $\mu\neq \pi$ this  a paraboloid opening towards the positive or negative $\gren$-direction, depending on the 
sign of $\mu$ (non-zero by assumptions).
When $\mu=\pi$  we obtain a plane in the three-dimensional $(\gren,\grp)$ space. Pictures of all these conjugacy classes can be found in \cite{FoFS}.

The planar   conjugacy class is  only one  which looks qualitatively different from any of the adjoint orbits obtained in Sect.~\ref{adorbsec}. The planarity means that the energy parameter $\gren$ does not depend on the momentum $\grp$, as confirmed by the dispersion relation in \eqref{uconj1}.  However, this phenomenon is not as strange as it may at first appear. A similar effect occurs  when one parametrises conjugacy classes in the group $SL(2,\RR)$ (which is the relativistic analogue of $\NW$) in the form $\exp(\gren J_0)\exp(-\grp\cdot \vec{J})$, where $J_0,J_1,J_2$ are the usual Lorentz generators of $SL(2,\RR)$.  When plotting the  conjugacy classes containing the element $\exp(\mu J_0)$ in the  $(\gren,\grp)$ space, they curve up or down, or are horizontal depending on the value of $\mu$, just like for $\NW$.

The centraliser group associated to the conjugacy classes \eqref{uconj1} is  the group $N_{E_0,m}
\simeq U(1)\times \RR$ already encountered in the discussion of adjoint orbits \eqref{paracent}.
The relation \eqref{grouplie} to the Lie algebra parameters  is 
\bea
\grp =\frac{\sin\frac \mu 2}{\frac m 2}\vec{p}, \qquad (\gren-\gren_0)=\frac{\sin \mu}{m}(E-E_0).
\eea

Finally, the  conjugacy class containing $u_0(\grp_0)$
 is 
\bea
\label{uconj2}
C_{\grp_0}=\{(0,\grp,\gren)||\grp| =|\grp_0|, \gren \in \RR\},
\eea
which  is a cylinder (see again \cite{FoFS} for a picture). The centraliser group is the group $N_{\vec{p}_0}\simeq \RR^2$
of \eqref{cylcent}. The relation \eqref{grouplie} to the Lie algebra parameters  is 
\bea
\grp =8\pi G\vec{p}, \qquad (\gren-\gren_0)=8\pi G(E-E_0).
\eea

Turning now to the conjugacy classes in the inhomogeneous group $\INW$, we would like  to parametrise the  holonomies \eqref{inhomoholo} in the form
 \bea
 \label{poispara}
 h=u\cdot \rel,
 \eea
where $u\in \NW$ and $r\in \INW$ is entirely inside the normal abelian subgroup, i.e. can be written
\bea
\label{relpara}
\rel = (\theta\sigma, -\theta\vec{\iota}, \theta \iota), \quad \sigma,\iota\in \RR, \vec{\iota}\in \RR^2.
\eea
The reason for this parametrisation is that  the Poisson structure, to be studied further below,
 takes a particularly simple form in these coordinates. Note that the
 factorisation is  similar to the semi-direct product factorisation
 \eqref{facor}, but of the opposite order.

Writing $u$ again as in \eqref{homopara} we multiply out to find
 \bea
 h=(\mu +\theta\sigma, -(\grp + \theta R(\mu)\vec{\iota}), \gren + \theta(\iota +\frac 1 2 \grp\times R(\mu)\vec{\iota}). 
 \eea
 In order find how the coordinates $\iota, \vec{\iota}$ transform under conjugation ($\sigma $ is invariant), we use the  following conjugation formula, with the conjugating element $g\in \INW$ 
 written in the semi-direct product parametrisation \eqref{facor} i.e. $g=xv$:
 \begin{align}
u' r'& = ( x  v) (u  r) (x  v)^{-1} \nonumber \\
 &= x  (v u v^{-1} \; v r v^{-1})
x^{-1} \nonumber \\
&= (vuv^{-1}) \; (vu^{-1}v^{-1} ~ x ~ vuv^{-1} ~v r v^{-1} ~x^{-1}).
\end{align}
or, in a more compact form
\bea
\label{conformula}
u'=vuv^{-1}, \qquad r' = (u')^{-1} x u' ~ x^{-1} ~ vrv^{-1},
\eea
where we have used the commutativity of the inhomogeneous part of $\INW$. Note that the multiplication of purely inhomogeneous elements, with all parameters proportional to $\theta$,
amounts to the {\em addition} of the corresponding parameters. 
 
 Using the general formula \eqref{conformula} we classify the
 conjugacy classes of $\INW$. The results are summarised in Table~\ref{inawiclasses}, which should be seen as the group analogue of the adjoint orbits in  Table~\ref{inawiorbits}. The two-dimensional orbits are, in fact, the same as in the Lie algebra case, except that they now live in the inhomogeneous, abelian part of the group $\INW$.
 The four-dimensional conjugacy classes  are  obtained by exponentiating the four-dimensional adjoint orbits into the group. Thus their topology is exactly the same as that of the corresponding adjoint orbits, but their embedding in the group, and hence their parametrisation in terms of coordinates \eqref{homopara} and \eqref{relpara}  is new. In particular the 'dispersion relations' between the group parameters defining  each classes take an unfamiliar form. We list them in each case. We also give
 reference  group elements from which the conjugacy classes can be obtained by conjugation.  The parameters appearing in them are related to those in the reference elements of  Table~\ref{inawiorbits} via \eqref{homorefs} and 
\begin{equation}
  \label{scaleparaa}
 \g{s} =8\g{p}G \sigma,\quad 
 \vec{\g{i}}_0=8\g{p}G \vec{j},\quad
  \iota_0=8\g{p}G \anj_0.
\end{equation}

\begin{table}[h! b! p!]
\baselineskip 30 pt
\begin{center}
\begin{tabular}{|c| p{6.5cm}| p{4.1cm} p{0.01cm}|}
\hline
%  &    &   \\
\centering \textbf{Reference element}

   &  \centering \textbf{Generic element in conjugacy class}

 &
 \centering \textbf{Geometry of conjugacy class}

&  \\
% constant & & \\
 % &    &   \\
\hline
%  &    &   \\
 \centering $(0, \vec{0}, \gren_0) (0,\vec{0},\theta \iota_0) $  & \centering $ (0, \vec{0}, \gren_0)(0,
\vec{0}, \theta \iota_0)~$ & \centering Point & \\
% &    &   \\
\hline
%  &    &   \\
 \centering $(0, \vec{0}, \gren_0) (\theta \sigma, \vec{0},  \theta \iota_0) $  & \centering $ (0, \vec{0}, \gren_0)(\g{8} \g{s},-
\g{8}\vec{\g{i}}, \theta \iota)~$ \newline with $ \vec{\g{i}} \in
\RR^2, \g{i} = \g{i}_0 +
 \tfrac{1}{2\g{s}}\vec{\g{i}}^{2} $ & \centering Paraboloid $ \simeq \RR^2 $ & \\
\hline
%  &    &   \\
 \centering $ (0, \vec{0}, \gren_0)(0,
\g{8}\vec{\g{i}}_0, \theta \iota_0) 
  $  & \centering $ (0, \vec{0}, \gren_0)(0,
-\g{8}\vec{\g{i}}, \theta \iota)~$\newline with $ 
\g{i} \in \RR, |\vec{\g{i}}| = |\vec{\g{i}}_0| $ & \centering Cylinder $ \simeq S^1 \times \RR $ & \\

\hline
 % &    &   \\
$(\mu, \vec{0}, \gren_0)(\theta \g{s},
 \vec{0}, \theta \iota_0)$
& \centering
$  (\mu, -\vec{\pi}, \gren)(\theta \g{s},
-\theta \vec{\iota}, \theta \iota) $ \newline with $ \gren = \gren_0 +
\dfrac{1}{4}\cot{\tfrac{\mu}{2}} \vec{\pi}^2, $ 
$ \iota
=\iota_0 + \frac{ \sigma \vec{\pi}^2}
{ 8  \sin^2 \tfrac{\mu}{2} }+\frac 1 2 (\epsilon  +\cot
\tfrac{\mu }{2})\vec{\pi}\cdot R(\mu)\vec{\iota} $ &\centering Tangent
bundle of paraboloid or plane $ \simeq \RR^4 $  &\\
 % &    &   \\

\hline
 % &    &   \\
\centering $  (0, -\vec{\g{p}}_0, \gren_0) (\theta\sigma , -\theta\vec{\g{i}}_0, \theta
\iota_0) $ &
\centering $ (0, -\vec{\g{p}}, \gren) (\theta\sigma , -\theta\vec{\g{i}}, \theta
\iota) $ \newline with $ |\vec{\g{p}}| = |\vec{\g{p}}_0|, \newline
\vec{\g{p}} \cdot \vec{\g{i}} -\gren \sigma = \vec{\g{p}}_0\cdot \vec{\g{i}}_0 -\gren_0\sigma
$ &\centering Tangent
bundle of cylinder $ \simeq S^1\times\RR^3 $  &\\
 % &    &   \\
\hline

\end{tabular}
\end{center}
\caption{ \label{inawiclasses} Conjugacy classes of the centrally extended Galilei group $\INW$}
\end{table}

\subsection{The phase space of Galilean 3d gravity and the Poisson-Lie structure of $\INW$}

The  phase space of Chern-Simons theory with gauge group $H$ on a three-manifold of the form
$\RR \times \Sigma $ is the space of flat $H$ connections on $\Sigma$, equipped with Atiyah-Bott symplectic structure \cite{AB,Atiyah}. This space can be parametrised in terms of holonomies  around the non-contractible loops in $\Sigma$, starting and ending at an arbitrarily chosen reference point,  modulo conjugation (the residual gauge  freedom at the reference point). For a very brief review of these facts in the context of 3d gravity we refer the reader to the talks \cite{Schroers,Schroerstalk},
where additional references can be found.   
The symplectic structure on the phase space can be described in a number of ways. The one which is ideally suited to the combinatorial quantisation approach used in this paper is a description due to Fock and Rosly \cite{FR}.  The idea is to work on the extended phase space of all holonomies around non-contractible loops on $\Sigma$ (without division by conjugation), and  to define a Poisson bracket on the extended phase space in terms of a classical $r$-matrix which solves the classical Yang-Baxter equation and which is compatible with the  inner product used in the Chern-Simons action.  The compatibility conditions is simply that the symmetric part of the  $r$-matrix equals the Casimir element associated with the inner product used in defining the Chern-Simons action.  
%The physical phase space is then obtained as finite-dimensional  quotient of this extended phase space. 

The classical $r$-matrix endows the gauge group with a Poisson structure which is compatible with its Lie group structure, i.e.  with a Poisson-Lie structure. It turns out that the Poisson structure on the extended phase space considered by Fock and Rosly can be described in terms of two standard Poisson structures associated to the Poisson-Lie group $PL$: one is  called the Heisenberg double (a non-linear version of the cotangent bundle of $PL$) and the other is the dual Poisson-Lie structure (a non-linear version of the dual of the Lie algebra of $PL$) \cite{AMII}.  The extended phase space for Chern-Simons theory on 
$\RR \times \Sigma_{gn}$, where $\Sigma_{gn}$ is a surface of genus $g$ with $n$ punctures, 
can then shown  \cite{AMII} to be isomorphic, as a Poisson manifold, to a cartesian product of 
$g$ copies of the  Heisenberg double of $PL$   and a   symplectic leaf of the  dual Poisson Lie group for every puncture.  The Fock-Rosly Poisson structure on the extended phase space for gauge groups of the semi-direct product form $PL=H\ltimes \mathfrak{h}^*$  is studied in detail in \cite{MS2}. Since the group $\INW$ is of this form, all of the results from 
\cite{MS2} can be directly imported. In particular, it is shown in \cite{MS2} that the Heisenberg double for groups of the form $H\ltimes \mathfrak{h}^*$  is simply the cotangent bundle of $H\times H$. The dual Poisson structure is more interesting. We now work it out in detail for the group $\INW$.

The Lie algebra $ \inawi $ has the structure of a classical double:  it is the direct
sum of the Lie algebras $ \nawi $  (spanned by $J,J_1,J_2,S$) and the abelian Lie algebra  spanned by $H,P_1,P_2,M$. These two Lie algebras are in duality via the pairing $\langle,\;,\;\rangle$,
so we can identify the abelian Lie algebra with the dual vector space $ \nawi^* $.  Thus $\inawi=\nawi\oplus\inawi^*$, and the data we have just described is precisely that of a Manin triple \cite{CP}.  It follows that $\inawi$ has, in fact, the structure of a Lie bi-algebra, which means that it has both commutators and co-commutators, with a compatibility between the two. The co-commutators can be expressed in terms of the 
$ r$-matrix
\begin{equation}
  \label{eq:r_matrices}
  r = M \otimes J+H\otimes S - P_i \otimes J_i.
\end{equation}
which pairs generators with their dual, and is clearly compatible with the inner product $\langle \;,\;\rangle$  in the sense of Fock and Rosly: the symmetric part of $r$ is precisely the Casimir \eqref{inwcasimir}.
On the classical double $ \inawi = \nawi\oplus\nawi^* $ the Lie
algebra structure is that of \eqref{extgal}, while the co-commutator it is given via
\begin{equation}
  \g{d}(X) = (\ad_X\otimes 1 + 1 \otimes \ad_X)(r), \qquad X\in \inawi \end{equation}
or explicitly
\begin{align}
  \label{eq:cocom_inawi}
  &\g{d}(S) = \g{d}(J) = 0, && \g{d}(J_i) = 0 \nonumber \\
&\g{d}(M) =  0,\; \g{d}(H) = -\epsilon_{ij}P_i\otimes P_j && \g{d}(P_i) = \g{e}_{ij}( M \otimes P_j - P_j \otimes M). 
\end{align}
As mentioned above, the $r$-matrix also gives rise to a Poisson structure on the group $\INW$, and to associated dual  and Heisenberg double Poisson structures. For explicit formulae in a more general context we refer the reader to \cite{MS2}. Here we note that the symplectic leaves of the dual Poisson structure (which are associated to the punctures in the Fock-Rosly construction) can be mapped to conjugacy classes in the original group $\INW$. The symplectic potentials for the symplectic  form  on these leaves can be expressed very conveniently  in manner which is entirely analogous  to the expression \eqref{adpotential} for the symplectic potentials  on (co)adjoint orbits. The only difference is that the  inhomogeneous Lie algebra element in \eqref{adpotential} is replaced by its group analogue in the factorisation \eqref{relpara}, leading to the symplectic potential
\bea
\label{dualsympl}
\tilde{\boldsymbol{\g{8}}}_{\INW^*}=\langle  (\sigma H -\vec{\iota}\cdot \vec{P} +\iota M) , \dd v v^{-1}\rangle,
\eea
where  $r=(\theta\sigma, -\theta\vec{\iota},\theta\iota)$  is  the
inhomogeneous part of a group element in the given conjugacy
class. By inserting the    parameters $\sigma, \iota, \vec{\iota}$ for the various conjugacy classes given in 
Table~\ref{inawiclasses} one obtains the symplectic potentials on each of the conjugacy classes listed there. One can attempt to simplify the resulting expressions using \eqref{conformula}, but we have not been able to generate illuminating formulae in this way.
% Note $  \dd v v^{-1} $ is
% \begin{equation}
%   \label{eq:right_inv_maur_cart}
%   \dd v v^{-1} & = (\dd\vec{w} -\dd\varphi \epsilon \vec{w} )\cdot \vec{J} + \dd \g{ff} J + (\dd
% \sig  +\frac 1 2 \vec{w}^2 \dd\varphi- \dfrac{1}{2} \dd\vec{w} \times \vec{w}) S
% \end{equation}
% so
% \begin{equation}
%   \label{eq:sympl_pot_}
%   \dd v v^{-1} & = (\dd\vec{w} -\dd\varphi \epsilon \vec{w} )\cdot \vec{J} + \dd \g{ff} J + (\dd
% \sig  +\frac 1 2 \vec{w}^2 \dd\varphi- \dfrac{1}{2} \dd\vec{w} \times \vec{w}) S
% \end{equation}

The upshot of the Fock-Rosly construction for Chern-Simons theory with gauge group $\INW$
is thus as follows.  For a universe of topology 
$\RR \times \Sigma_{gn}$,  and assuming  for definiteness that the punctures are decorated with  coadjoint orbits $O_{m_i,s_i}$, $i=1\ldots n$, representing particles with masses $m_i$ and spins $s_i$,  
 the extended phase space is 
\bea 
\label{extspace}
\tilde {\cal P} = \INW^{2g}\times { C}_{\mu_n \sigma_n}\times \ldots {C}_{\mu_1 \sigma_1},
\eea
where we have written $C_{\mu_i\sigma_i}$ for the conjugacy classes obtained by exponentiating the orbits $O_{m_is_i}$, and assumed that none of the $\mu_i$ equal an integer multiple of $2\pi$.  We have also suppressed the labels $E_0,j_0$ and their group analogues $\gren_0,\iota_0$ since they do not affect the orbit geometry and symplectic structure. 

The physical  phase space is the quotient
\begin{align}
\label{pspace}
 {\cal P} & = \{(A_g,B_g,\ldots, A_1,B_1, M_n,\ldots M_1)\in \tilde {\cal P}|\nonumber  \\
   & \quad [A_g,B_g^{-1}]\ldots[ A_1,B_1^{-1}]M_n\ldots M_1=1\}/\mbox{conjugation}.
\end{align}
The non-trivial part of the phase space, apart from the quotient, are the conjugacy classes in  $\INW$. As explained above, these are precisely the symplectic leaves of the dual Poisson-Lie group  $\INW^*$, with symplectic potential \eqref{dualsympl}.

\section{Towards Galilean quantum gravity in  2+1  dimensions}
\label{quantumsection}

Our description of classical Galilean gravity in terms of a Chern-Simons theory and a compatible $r$-matrix   is tailor-made for the application of the  combinatorial quantisation programme summarised in the introduction. As explained there,  all aspects of the quantisation  - the construction of the Hilbert space, the representation of observables and the implementation of  symmetries on the Hilbert space - are determined by a  quantum group associated to the gauge group used in the Chern-Simons action and the compatible $r$-matrix. Since the centrally extended Galilei group with the 
classical $r$-matrix \eqref{eq:r_matrices} is an example of a classical double of the form 
$H\ltimes \mathfrak{h}^*$, it follows from the general results of  \cite{MS2}  that the relevant quantum group for Galilean quantum gravity is the quantum double  $D(\NW)$  of the   homogeneous part of the centrally extended Galilei group or, in short,  the Galilei double.
The focus of this final section is therefore the definition and  representation theory of $D(\NW)$. Since this is closely related to the representation theory of  centrally extended Galilei group $\INW$ itself, we begin with a quick review of the latter.

\subsection{Representation theory of the centrally extended Galilei group}

The particular version of the centrally extended Galilei group studied in this paper is one of 
the central extensions of the Galilei group whose representation theory is studied in \cite{Grigore}.
We briefly review  the representation theory in our notation, with a particular emphasis on the fact that, for our version of the extended Galilei group, the 
representation theory easily follows from the general representation theory of semi-direct products as developed in \cite{Mackey} and explained, for example,  in \cite{BaRa}. 

Irreducible representations of semi-direct products of form $H\ltimes N$,  where $N$ is an abelian group, are labelled by $H$-orbits on the dual of $H$ (the space of characters) and irreps  of centraliser groups of  chosen  points on those orbits. In the case at hand, where the semi-direct product is of the particular form $\NW\ltimes \nawi^*$, the dual of the abelian group $\nawi^*$ is the Lie algebra $\nawi$ itself (viewed as a vector space). The relevant orbits in this case are therefore  the adjoint orbits of $\NW$ which we listed, together with associated centraliser groups, in Sect.~\ref{adorbsec}.

 As for general semi-direct products,   we can describe the irreps either in terms of functions on
 the homogeneous group $\NW$ obeying an equivariance condition, or in terms of functions directly on the  orbits  (this is rather analogous to the description of the symplectic leaves in Sect.~\ref{poisssect} either in terms of the conjugating group element  or the orbit itself).   The description in terms of functions on  $\NW$ has a simpler formula for the action of $\INW$ elements, and easily generalises to the quantum group $\INW$.  However, in the literature on the Galilei group, the second view point is commonly used, and we will adopt it in this paper, too. Since the representation theory is well-documented in the literature \cite{Grigore,Mund}, we only sketch the method and then  illustrate it with one case in order to establish a clear dictionary between our notation and that used, for example, in \cite{Grigore}

 For definiteness we consider the case  of a massive particle with spin. The relevant adjoint orbit of 
 $\NW$ is the orbit $O_{E_0,m}$ \eqref{paraorbit}, which we can parametrise entirely in terms of the unconstrained momentum vector $\vec{p}$, with the energy given by
 \bea
 \label{disprel}
E=E_0+ \frac {1} {2m}\vec{p}^2.
\eea
Homogeneous elements $v\in \NW$ act on  the  orbit via adjoint action $\Ad_v$ \eqref{adnw} and, by a slight abuse of notation, we write this action as  $\Ad_v(\vec{p})$. In other words, it is understood that $E$ changes under  Galilei boosts according to \eqref{adnw}, which ensure that the energy-momentum relation  \eqref{disprel} is maintained. The carrier space of the irreducible representation is then simply the space $L^2(\RR^2)$ of square-integrable functions on the space of momenta. 

In order to write down the action of a general element  $g\in \INW$ on states $\psi\in L^2(\RR^2)$ we need the reference point $Q_0(E_0,m)$ \eqref{qem} on the orbit, and we need to pick a map
\bea
\label{sectionmap}
S: O_{E_0,m}\rightarrow \NW,
\eea
which associates to a given point $Q=mJ-\vec{p}\cdot\vec{J} + ES  \in  O_{E_0,m}$
a  $\NW$-element  $S(\vec{p})=v$ satisfying
\bea
\Ad_v(Q_0)=Q,
\eea
i.e. $v$ is an element of the homogeneous Galilei group which rotates/boosts the reference point $Q_0$  to the given point $Q$  on the adjoint orbit.  
 In the case at hand, we can, for example, pick
 \bea
 S(\vec{p}) = (0,-\epsilon\frac {\vec{p}}{m},0),
 \eea
where we have again used $\vec{p}$ as a coordinate on $O_{E_0,m}$.
Then we define the so-called multiplier  function  \cite{BaRa} 
\bea
\label{multiplierphase}
n:  \NW\times O_{E_0,m}\rightarrow N_{E_0,m},\qquad 
  n(v,\vec{p})=(S(\vec{p}))^{-1}v \, S(\Ad_{v^{-1}}(\vec{p})). 
\eea
The key properties of this function (both easy to check) are  that it takes values in the centraliser group $N_{E_0,m}$ \eqref{paracent} associated to the orbit $ O_{E_0,m}$,  and that it satisfies the cocycle condition
\[
n(v_1,\vec{p})n(v_2,\Ad_{v_1^{-1}}\vec{p})=n(v_1v_2,\vec{p}).
\]
Computing explicitly, we find, for $v=(\varphi, \vec{w}, \sig)$, again with the convention $\vec{v}=-\epsilon \vec{w} $,  that
\bea
\label{masmult}
n(v,\vec{p}) = (\varphi, \vec{0}, \sig + \frac {1} {2 m} \vec{p}\times \vec{v}).
\eea
Then, picking an irrep $\pi_{\anj_0,s}$  of the centraliser group $N_{E_0,m}\simeq
U(1)\times \RR$, labelled by an integer $\anj_0 $ and a real number $s$, 
the action of an  element  $g=(\theta\alpha, \theta\vec{a}, \theta \tau)(\varphi,\vec{w},\sig)\in \INW$  parametrised in the semi-direct product form \eqref{facor} on a state
 $\psi\in L^2(\RR^2)$ is 
\begin{align}
\label{massiverep}
\left( \Pi_{E_0,m,\anj_0, s}(g)\;\psi\right)(\vec{p})&=\pi_{\anj_0,s}(n(v,\vec{p}))e^{i(m\tau + \vec{a}\cdot\vec{p} +E \alpha)}\;\psi(R(-\varphi)(
\vec{p}-m\vec{v})) \nonumber  \\
&=e^{i(\anj_0\varphi+ s(\sig +
\frac {1}{ 2m} \vec{p}\times\vec{v})+ m\tau + \vec{a}\cdot\vec{p} + \alpha(E_0+ \frac{1}{2m}\vec{p}^2)}\;\psi(R(-\varphi)(
\vec{p}-m\vec{v})).
\end{align}
In comparing this expression with representations in the literature readers should be aware of the different parametrisation \eqref{gal_group} frequently used for the Galilei group, and the relation to our semi-direct parametrisation explained after \eqref{gal_group}.

It is not difficult to adapt this example to the representations corresponding to the cylindrical orbits \eqref{cylorbit}. The centraliser group is  $N_{\vec{p}_0}\simeq \RR\times \RR$, so irreps are labelled by the orbit label $\vec{p}_0$ and two real numbers $s,k$ characterising an irrep $\RR\times \RR$. Finally, note that irreps of $\INW$ corresponding to the vacuum orbit $\eqref{qe}$ are irreps of the group $\NW$,  which were studied in \cite{KK}

\subsection{The Galilei double and its irreducible representations}

The transition from the inhomogeneous group $\INW$ to the quantum double $D(\NW)$ can
be most easily understood at the level of representations \cite{Schroers}. Looking at \eqref{massiverep} we see that the translation part $(\theta\alpha,\theta \vec{a}, \theta \tau)$ acts on 
states by multiplication with the function $\exp(i(\alpha E  +\vec{a}\cdot\vec{p} +  m\tau))$ on 
momentum space $\nawi$. In the quantum double, these 'plane waves' are replaced by general functions on the {\em group} $\NW$. At the purely algebraic level, this
 makes   no difference, but the non-abelian nature of $\NW$ makes a crucial difference for the 
 co-product structure. The quantum double $D(\NW)$, called Galilei double in this paper,  is a non-co-commutative Hopf algebra, with a non-trivial $R$-matrix and ribbon element.

 As an algebra, the quantum double of a group $H$ is the semi-direct product of the group algebra of $H$ with algebra of functions on $H$.  It is not trivial to give a mathematically rigorous definition of this for the case of  a    Lie group $H$, and we refer the reader to \cite{KM}
where the quantum double is identified with continuous
functions on $H\times H$. Further details and the application to three-dimensional quantum gravity are discussed in \cite{BM} and \cite{Schroers, BMS}. For a full list of the Hopf algebra structure in our conventions, including antipode,  co-unit and $*$-structure,  we refer the reader to \cite{MS2}.  Here we adopt a naive approach where elements of $D(H)$ are  described as a tensor product of a group element $v\in H$ and a function $f$ on $H$.  Such elements correspond to singular elements in $D(H)$ in the definition of \cite{KM}, but they are convenient for 
summarising the  algebra and co-algebra structure. 
\begin{align}
\label{singdoublemult}
&(v_1\otimes  f_1)\bullet (v_2 \otimes f_2)
= v_1v_2 \otimes ( f_1\cdot f_2\circ\Ad_{v_1^{-1}})\nonumber \\
&\Delta(v\otimes f)(u_1, u_2)=v \otimes  v\; f(u_1u_2).
\end{align}
The universal
$R$-matrix of  $D(H)$ is an element of $D(H)\otimes D(H)$;   in our singular notation it is
\bea
\label{Rmat}
R = \int_H \; dv (e\otimes \delta_v)\otimes (v\otimes 1), 
\eea
where $e$ stands for the identity element of $H$ and  $1$ for the function on $H$ which is  $1$ everywhere.  

 The   irreps of quantum doubles of a continuous Lie group $H$, classified in \cite{KM},  are labelled by the
  $H$-conjugacy classes and irreps of associated centraliser groups. For the case of the extended homogeneous Galilei group $\NW$, this is precisely the data  we collected in equations \eqref{uconj1} to \eqref{uconj2}. The study of the representation theory in \cite{KM}
   is in terms of equivariant functions on the group, but it is not difficult to express the irreps in
 terms of functions on the conjugacy classes and multipliers like  \eqref{masmult}. In fact, this is language employed for irreps of the double of a finite group in \cite{BDWP}. 
 
 Consider for definiteness the irreps corresponding to the conjugacy classes $C_{\gren_0,\mu}$ \eqref{uconj1},  with associated centraliser group $N_{E_0,m}\simeq U(1)\times \RR$  \eqref{paracent}.  Analogously to the adjoint orbits $O_{E_0,m}$ the conjugacy classes can be parametrised by the  rescaled momentum vectors $\grp$, with the rescaled energy determined by
 \bea
 \label{disprel2}
\gren=\gren_0 +\frac 1 4 \cot\left(\frac \mu 2\right)\grp^2.
\eea
The carrier space of the irrep  is  the space of square-integrable functions on the class
 $C_{\gren_0,\mu}$.
Using the parametrisation of that orbit in terms of the momentum $\grp$ we could again identify the carrier space with  $L^2(\RR^2)$,  but this is less convenient in this case. Instead we
use the parametrisation \eqref{homopara} for elements $u\in C_{\gren_0,\mu}$, where the relation
\eqref{disprel2} is understood.  With $u_0(\mu, \epsilon_0)$ as in \eqref{homorefs} we have, from \eqref{ccformula}, 
\bea
\label{groupboost}
\Ad_v(u_0(\epsilon_0,\mu))=(\mu,1-R(\mu)\vec{w}, \gren_0+\frac 1 2 \sin^2\mu \vec{w}^2).
\eea
The analogue of the map \eqref{sectionmap} is now a map
\bea
S: C_{\gren_0,\mu}\rightarrow \NW,
\eea
which associates to $u\in C_{\gren_0,\mu}$ an element $v=S(u)$ so that 
\bea
\Ad_v(u_0(\epsilon_0,\mu))= u.
\eea
Using \eqref{groupboost} we can pick, for example,
\bea
S(u) = (0,(R(\mu)-1)^{-1}\grp,0),
\eea
so that the analogue of  multiplier function \eqref{masmult} is now
\bea
\label{gmasmult}
n(v,u) = (\varphi, \vec{0}, \sig + \frac {1} {2}(R(\mu)-1)^{-1}\grp \cdot\vec{v}).
\eea
Finally we again pick an irrep $\pi_{\iota_0 \sigma}$ of the centraliser group $N_{E_0,m}\simeq
U(1)\times \RR$, labelled by an integer $\iota_0 $ and a real number $\sigma$,  and denote the 
carrier space $L^2(C_{\gren_0,\mu})$ of the associated irrep of $D(\NW)$   by $V_{\gren_0,\mu,\iota_0,\sigma}$.
 Then singular elements $v\otimes f\in D(\NW)$  act  on  $\psi \in V_{\gren_0,\mu,\iota_0,\sigma} $
 via
\begin{align}
\label{gmassiverep}
\left( \Pi_{\gren_0,\mu,\iota_0,\sigma}(v\otimes f)\;\psi\right)(u)&=\pi_{\iota_0,\sigma}(n(v,u)) f(u)\;\psi(\Ad_{v^{-1}}(u)) \nonumber \\
&=e^{i(\iota_0\varphi+ s(\sig +
\frac {1}{ 2}(R(\mu)-1)^{-1}\grp \cdot\vec{v})} f(u)\;\psi(\Ad_{v^{-1}}(u)).
\end{align}

Analogous formulae can be derived for the other conjugacy classes. 
For the  trivial conjugacy classes $C_{\gren_0}$ , the centraliser  group  is the entire group $\NW$. As in the case of the centrally extended Galilei group, the irreps of $\NW$
again appear as special irreps of $D(\NW)$.
The  irreps of $D(\NW)$ labelled by conjugacy classes  of the form $C_{\grp_0}$ 
correspond to massless particles, and closely resemble the corresponding irreps of $\INW$.

\subsection{Aspects of Galilean quantum gravity in 2+1 dimensions}

As reviewed at the beginning of this section, the combinatorial quantisation programme  expresses all aspects of the quantisation of  the phase space  \eqref{pspace} in terms  of the representation theory of $D(\NW)$. Having studied this representation theory, we end this paper by highlighting some of the features of the quantum theory which can be derived without much further calculation. 
According to \cite{MS2}, 
 the Hilbert space obtained by quantising the phase space \eqref{pspace} for $n$ massive and spinning particles on a surface $\Sigma_{gn}$ of genus $g$ is 
 \bea
 \label{Hilbertius}
{\cal H} = \text{Inv}\left(L^2(\NW^{2g})\; \otimes \;\bigotimes_{i=1}^n V_{\gren_{0,i},\mu_i,\iota_{0,i},\sigma_i} \right),
\eea 
where Inv means the invariant part of the tensor product under the action of $D(\NW)$. 
The general structure of this formula is not difficult to understand. Each Heisenberg double (the Poisson manifold associated to the handles, which is the cotangent bundle of $\NW\times \NW$ in the our case)  quantises to $L^2(\NW\times \NW)$, and each conjugacy class (associated to punctures) quantises to an irrep of $D(\NW)$. The division by conjugation that takes one from the extended phase space \eqref{extspace} to the physical phase space \eqref{pspace} is mirrored in the quantum theory by the restriction to the invariant part of the tensor product in \eqref{Hilbertius}.
We refer the reader to \cite{MS2} for  further  details, and focus on qualitative aspects here.

\subsubsection*{Fusion rules}
 
 In order to pick out the invariant part of the tensor product in
 \eqref{Hilbertius} one needs to be able to decompose tensor products
 of representations of $D(\NW)$ into irreps.  A general method for
 doing this in the case of  quantum doubles of compact Lie groups is
 explained in \cite{KMB}, where it is worked out in detail for $D(SU(2))$.   Extending this to include the case of $D(\NW)$ is  a technical challenge, but  certain aspects of the tensor product decomposition can be read off without much work.

 The tensor product decomposition of   products like
 \bea
  V_{\gren_{0,1},\mu_1,\iota_{0,1},\sigma_1}\otimes  V_{\gren_{0,2},\mu_2,\iota_{0,2}\sigma_2},
 \eea
 which occur in \eqref{Hilbertius},
  corresponds physically to a set of fusion rules
for particles in Galilean gravity. They tell us how to 'add' kinetic attributes like energy and momentum of two particles. In the  representation theory of the Galilei double $D(\NW)$,  the conjugacy classes labelling irreps are  combined by multiplying out the group elements contained in them, and sorting the products into conjugacy classes again. The underlying rule for combining 
energy and momentum is thus simply to multiply out the corresponding $\NW$ elements\footnote{This fact can be derived more generally for Chern-Simons descriptions of particles in 2+1 dimensions from the fact that holonomies get multiplied when the underlying loops are concatenated, see e.g. \cite{BM}}. Thus, with
\bea
\label{twops}
u_1=(\mu_1,-\grp_1,\gren_1),\quad u_2=(\mu_2,-\grp_2,\gren_2)
\eea
we have 
\bea
 u_1u_2 =(\mu_1+\mu_2,
   -(\grp_1 + R(\mu_1)\grp_2),\gren_1+\gren_2 +\frac 1 2 \grp_1\times R(\mu_1)\grp_2).  
   \eea
This simple rule has  an interesting consequence.  Even if $u_1$ and $u_2$ both belong to conjugacy classes of the type $C_{E_0,\mu}$ \eqref{uconj1} describing massive particles,  it is possible for the product  $u_1u_2$ to belong to belong to a conjugacy class of the type $C_{\grp_0}$ \eqref{uconj2} describing massless particles. This happens generically
when the sum of the particle's (rescaled) masses is $2\pi$, which  essentially means that, in physical units,  the total mass equals the Planck mass.  Such particles are the non-relativistic analogue of the so-called Gott pairs in relativistic 3d gravity \cite{Gott}. However, unlike in the relativistic case there is no condition on the relative speed of the two particles. As long as the sum of the (rescaled) masses is $2\pi$, the total energy and momentum is either of the vacuum type or, generically,  of the massless type.

\subsubsection*{Braid interactions}

The mapping class group of the spatial surface $\Sigma_{gn}$ acts on the Hilbert space \eqref{Hilbertius}. If punctures are present, this means in particular that the braid group on $n$ strands of $\Sigma_g$
acts. The action of the generators of the braid group   on the Hilbert space  can be expressed in terms of the $R$-matrix \eqref{Rmat} \cite{MS}, and have a simple action on the $\NW$-parts of the holonomies
 around any two punctures, which can again be derived from more elementary considerations, as explained in \cite{BM}.  Assuming that the two punctures correspond to massive particles we denote 
 their $\NW$-valued energy-momenta again by $u_1$ and $u_2$ as in \eqref{twops}. In the quantum theory we can describe states of definite energy and momentum in terms of delta-functions $\delta_{u_1}$ and $\delta_{u_2}$ on the conjugacy classes corresponding to the particles' masses. 
  Either by acting with the $R$-matrix  \eqref{Rmat} on the the tensor product $\delta_{u_1}\otimes  \delta_{u_2}$  or via the elementary considerations in \cite{BM}  one finds that the action of the braid group generator  is
\bea
(u_1,u_2)\mapsto (u_1 u_2u_1^{-1}, u_1),
\eea
which we evaluate, using \eqref{twops}:
\begin{align}
 \label{conjugationformula}
u_1 u_2u_1^{-1}=(\mu_2, -(1- R(\mu_2))\grp_1 -R(\mu_1)\grp_2, 
 \gren_2 +\frac 12(1+R(\mu_2))\grp_1\times R(\mu_1)\grp_2 +\frac 1 2  \sin\mu_2\grp_1^2).
 \end{align}
 The braid group action can be related to scattering of two massive particles \cite{BMS}, which was one of first aspects of 3d quantum gravity to be studied in detail in the limit where one particle  is heavy, and the other can be treated as a test particle moving on the background spacetime created by the heavy particle \cite{tHooft2,DJ}.  In usual 3d gravity, the scattering can be understood in terms of the test-particle's motion on the conical spacetime due to the heavy particle. The deficit angle of the conical spacetime is proportional to the mass of the heavy particle in Planck units, and this deficit angle controls the scattering cross section \cite{tHooft, DJ}. 
  We can recover aspects of this scattering from the formula
 \eqref{conjugationformula}.   Taking the first particle to be the heavy particle, and setting its momentum to zero (so that we are in the heavy particle's rest frame), the conjugation action on the second (test) particle's momentum is simply 
 \bea
 \grp_2\mapsto  R(\mu_1)\grp_2.
 \eea
This is precisely the distance-independent deflection of the momentum  direction by an angle proportional to the heavy particle's mass that is characteristic of scattering on a cone.  It would be interesting to use  the scattering formalism of \cite{BMS}, using the $R$-matrix \eqref{Rmat}, to investigate the non-relativistic scattering in more detail, and to compare with \cite{tHooft, DJ}.
 
\subsubsection*{ Non-commutative geometry}
 
 The appearance of a  non-commutative Lie group as a momentum space can be related to non-commutativity of spacetime coordinates using the representation theory of the quantum double \cite{BaMa}. The basic idea is to apply a non-abelian Fourier transform to switch from the momentum-space wave functions $\psi$ in \eqref{massiverep} to  position-space representation, with a wavefunction necessarily living on  a non-commutative spacetime and obeying a linear wave equation, see also \cite{MS}.  Applying the reasoning of \cite{BaMa} and \cite{MS} to the Galilei double, where momenta are coordinates (i.e. functions) on the group $\NW$, 
  one expects the relevant non-commutative space algebra in this case to be  the universal enveloping algebra of the  Lie algebra  $\nawi$,  with generators now interpreted as position coordinates $X_1,X_2$,  a time coordinate $T$ and an additional generator $R$:
 \bea
 \label{spacetimenon}
 [X_i,R]=8\pi G  \hbar \; \epsilon_{ij}X_j,\qquad [X_1,X_2]= 8\pi G  \hbar \; T .
 \eea
 Note that the combination $\hbar G$  has the dimension length$^2/$time, so cannot simply be interpreted as a Planck length or Planck time. This dimension is consistent with the interpretation of $X_i$ as positions and of $T$ as time.  
 
 The algebra \eqref{spacetimenon} has a a number of interesting features. The operator representing Galilean time is a central element, and in that sense Galilean time remains absolute in the non-commutative regime.  It is also worth noting that the algebra of spacetime coordinates closes, and is simply the universal enveloping algebra of the Heisenberg algebra.   For fixed time, the  non-commutativity 
 of the space coordinates is of Moyal-type, with the deformation matrix $\Theta_{ij}=\epsilon_{ij} 8\pi G  \hbar  t $ proportional to time $t$ (an eigenvalue of $T$).   Algebraically,  this is precisely the non-commutative position algebra which was studied  in the context of planar Galilean physics in \cite{LSZ,DH0,HP1,HP2}. However, the status here is different. Our non-commutativity  is a quantum gravity effect  (proportional to both $G$ and $\hbar$) and thus  universal.  The non-commutativity of position coordinates of a particle in \cite{LSZ} is related to the properties of a particular particle moving in a classical and commutative plane.  Presumably the two effects would have to be somehow combined if such a 
 particle was moving in the non-commutative spacetime \eqref{spacetimenon}.  
  It would clearly be interesting to study  the spacetime \eqref{spacetimenon} and its physical implications in detail.  
    The role of the generator $R$ (the only compact generator in $\nawi$) is not clear to us.  Its appearance reminiscent of  the appearance of a fourth dimension in the study of the  differential calculus \cite{FM} on the relativistic non-commutative spacetimes  with brackets $[X_a,X_b] =\epsilon_{abc}X^c$. 
 
 \vspace{0.2cm}
 {\bf Acknowledgements}
 
  BJS thanks DESY and the Perimeter Institute for hospitality during
  the time when this paper was written, and acknowledges a DAAD
  visiting fellowship for senior academics to fund the extended stay
  at DESY. GP acknowledges a PhD Scholarship by the Greek State
  Scholarship Foundation (I.K.Y).

\bibliographystyle{siam}

\begin{thebibliography}{99}
\bibitem{Christian} J.~Christian,  Exactly soluble sector of quantum gravity, Phy.~Rev.~D 56 (1997) 4844--4877; gr-qc/9701013.
\bibitem{Carlip} S.~Carlip, Quantum gravity in 2+1 dimensions,
  Cambridge University Press, Cambridge, 1998.
\bibitem{Grigore} D.~R.~Grigore,  The Projective unitary irreducible representations of the
                  Galilei group in (1+2)-dimensions, J.~Math.~Phys.,
                  37 (1996)  460-473; hep-th/9312048.
  % both Mund and Grigore talk about central extension of Galilei Lie
  % algebra - not group
 \bibitem{Mund} J.~Mund and R.~Schrader,  Hilbert spaces for
   nonrelativistic and relativistic 'Free' Plektons (Particles with
   braid group statistics), hep-th/9310054.
 \bibitem{BGGK}Y.~Brihaye, C.~Gonera,  S.~Giller and  P.~Kosinski, 
 Galilean invariance in (2+1)-dimensions,  hep-th/9503046.
 % describes three central extensions of Galilei Lie algebra
\bibitem{LevyLeblond} J.~M.~L\'evy-Leblond, in E.~Loebl (Ed.), Group Theory and Applications, Academic Press, New York, 1972.
\bibitem{Bose} S.~K.~Bose, The Galilean group in 2+1 space-times and its central extension,
  Commun.~Math.~Phys.~ 169 (1995) 385--395
  % this is an excellent paper, and clarifies that the Galilei group
  % has two central extensions, the Lie algebra has three
   \bibitem{BGdO} A.~Ballesteros, M.~Gadella and M.~A.~ Del Olmo,
 Moyal Quantisation of 2+1 dimensional Galilean systems, J.~Math.~Phys.~ 33 (1992) 3379-3386.
 % this paper is the best reference for coadjoint orbits and their quantisation and mentions relation of central extension to angular momentum
   \bibitem{LSZ} J.~Lukierski, P.~C.~Stichel and W.~J.~Zakrzewski,  Galilean invariant (2+1)-dimensional models with a Chern-Simons-like term and D = 2 noncommutative geometry.
 Annals Phys.~260 (1997) 224--249;  hep-th/9612017.
% explain how to obtain model with extended Galilei symmetry from a particle model with CS term
\bibitem{DH0}C.~Duval and P.~A.~Horvathy, The  'Peierls substitution' and the exotic Galilei group,  Phys.~Lett.~B 479 (2000) 284--290; hep-th/0002233.
%this seems to be the starting point of the extended discussion with Jackiw
\bibitem{HP1}  P.~A.~Horvathy and M.~S.~Plyushchay, Nonrelativistic anyons and exotic Galilean symmetry,  JHEP 0206:033 (2002);  hep-th/0201228.
%has clear discussion of the two kinds on non-commutative coordinates, and relation to Foldy-Wouthusyen and Newton-Wigner coordinates
\bibitem{HP2}  P.~A.~Horvathy and M.~S.~Plyushchay,   Anyon wave equations and the noncommutative plane, Phys.~Lett.~B595 (2004) 547--555; hep-th/0404137.
 \bibitem{JN} R.~Jackiw, and V.~P.~Nair, Anyon spin and the exotic central extension of the planar Galilei  group, Phys. Lett.  B 480 (2000) 237--238;
               hep-th/0003130. 
               % first all-out claim that second central relates to
               % spin 
 \bibitem{JN2}R.~Jackiw, and V.~P.~Nair,  Remarks on the  exotic central extension of the planar Galilei  group, Phys. Lett.  B 551 (2003) 166--168.
 %this paper points out the similarity between spin and rest mass,
 %clarifying discussion triggered by  their first paper
 \bibitem{NdOT} J. ~Negro, M.~A. ~del Olmo and  J. Tosiek, 
  Anyons, group theory and planar physics,  J.~Math.~Phys.~47 (2006) 033508; 
  math-ph/0512007. 
  % has coadjoint (some) orbits and coupling to electro-magnetic field
  % - for both Poincare and Galilei case
 \bibitem{DH}    C.~Duval  and  P.~A.~Horvathy,  Spin and exotic Galilean symmetry.
  Phys.~Lett.~B 547 (2002) 306--312;  Erratum-ibid.B588 (2004) 228.
 %argues that central extension is connected to non-commutativity of space and not necessarily to spin, HAS THE CONTRACTION FROM CENTRAL EXTENSION OF POINCARE
 \bibitem{Hagen} C.~R.~Hagen, Second central extension in Galilean covariant field theory, Phys.~Lett.~B 539 (2002) 168.
 %argues that the second central extension is not necessarily connected to spin
 \bibitem{BLL} 
H.~Bacry and  J. Levy-Leblond, Possible Kinematics, J.~Math.~Phys.~9 (1968) 1605--1614.
\bibitem{HS} F.~J.~Herranz and M.~Santander, 2+1 kinematical expansions: from Galilei to de Sitter algebras, math-ph/9902006.
 \bibitem{CJ}D.~Cangemi and R.~Jackiw, Gauge-invariant formulations of lineal gravities,
 Phys.~Rev.~Lett.  69 (1992) 233--236.
\bibitem{Nappi:1993ie} C.~R.~Nappi  and E.~Witten,  A WZW
  model based on a nonsemisimple group, Phys.~Rev.~Lett. 71 (1993)
  3751--3753; hep-th/9310112.
\bibitem{FoFS}
J.~Figueroa-O'Farrill and Sonia Stanciu,  More D-branes in the Nappi-Witten background,
JHEP (2000)  0001:024.
\bibitem{FR} V.~V.~Fock and A.~A.~Rosly, { Poisson structures on
moduli of flat connections on Riemann surfaces and $r$-matrices,} ITEP preprint { 72-92}  (1992);\newline
 V.~V.~Fock and A.~A.~Rosly, { math.QA/9802054}.
\bibitem{AGSI} A.~Y.~Alekseev, H.~Grosse and V.~Schomerus, {
Combinatorial quantization of the Hamiltonian Chern-Simons Theory,}
Commun.~Math.~Phys. { 172} (1995) 317--358.
\bibitem{AGSII} A.~Yu.~Alekseev, H.~Grosse and V.~Schomerus, {
Combinatorial quantization of the Hamiltonian Chern-Simons Theory
II,} Commun.~Math.~Phys. { 174} (1995) 561--604.
\bibitem{AS} A.~Yu.~Alekseev and V.~Schomerus, { Representation
theory of Chern-Simons observables,} Duke Math.~Journal { 85}
(1996) 447--510.
\bibitem{Schroers} B.~J.~Schroers, {Combinatorial quantisation of
Euclidean gravity in three  dimensions}, in:  N.~P.~Landsman,
M.~Pflaum, M.~Schlichenmaier (Eds.), {
Quantization of singular symplectic quotients}, Birkh\"auser, in:
Progress in Mathematics, Vol.~198,
 2001, 307--328; { math.qa/0006228}.
\bibitem{BNR} E.~Buffenoir, K.~Noui and P.~Roche,  Hamiltonian
Quantization of Chern-Simons theory with $SL(2,\CC)$ Group,
{Class.~Quant.~Grav.} {19} (2002)  4953-5016.
 \bibitem{BMS} F.~A.~Bais, N.~M.~Muller and B.~J.~Schroers, Quantum
 group symmetry and particle scattering in (2+1)-dimensional quantum gravity,
Nucl.~Phys.~B  640 (2002)  3--45
\bibitem{MS1} C.~Meusburger and B~J.~Schroers, Poisson structure and
  symmetry in the Chern-Simons formulation of (2+1)-dimensional
  gravity,
{ Class.~Quant.~Grav.} {20} (2003) 2193--2233.
\bibitem{MS2}  C.~Meusburger and B~J.~Schroers,
 The Quantization of Poisson structures arising in Chern-Simons theory with gauge group 
 $G\ltimes {\mathfrak g}^*$.
Adv.~Theor.~Math.~Phys.7 (2004) 1003--1043; hep-th/0310218.
\bibitem{Martin}S.~P.~Martin, Observables in 2+1 dimensional gravity, Nucl.~Phys.~B 327 (1989)  178--204.
\bibitem{Meusburger} C. ~Meusburger, Geometrical (2+1)-gravity and the Chern-Simons formulation: Grafting, Dehn twists, Wilson loop observables and the cosmological constant,
 Commun.~Math.~Phys.~273 (2007) 705--754; gr-qc/0607121.
\bibitem{MS6}C.~Meusburger and B.~J.~Schroers, Quaternionic and Poisson-Lie structures
in 3d gravity: the cosmological constant as a deformation  parameter,
J.~Math.~Phys.~49 (2008) 08351;
 arXiv:0708.1507.
\bibitem{Arnold} V.~I.~Arnold, Mathematical Methods of Classical
  Mechanics, Springer Verlag, New York, 1989.
 \bibitem{Kirillov} A.~A.~Kirillov, {Elements of the theory of
    representations}, Grundlehren der mathematischen Wissenschaft 220,
Springer Verlag, Berlin, 1976.
\bibitem{AT} A.~Achucarro and P.~Townsend,  {A Chern--Simons
 action for three-dimensional anti-de Sitter supergravity
 theories}   Phys.~Lett. B  180 (1986) 85--100.
\bibitem{Witten} E. Witten, 2+1 dimensional gravity as an exactly
 soluble system,  Nucl.~Phys. B { 311} (1988)  46--78.
\bibitem{DJtH} S.~Deser, R.~Jackiw and G.~'t Hooft G, Three-dimensional
  Einstein gravity: dynamics of flat space,  Ann.~Phys. NY 152 (1984)
  220--235.
\bibitem{SousaGerbert} P.~de Sousa Gerbert,  On spin and (quantum)
gravity in 2+1 dimensions, Nucl.~Phys.~ B346 (1990) 440--472. 
\bibitem{MS4} C.~Meusburger and B.~J.~Schroers, Boundary conditions and symplectic structure in the Chern-Simons formulation of (2+1)-dimensional gravity, Class.~Quant.~Grav. 22 (2005) 3689--3724; gr-qc/0505071.
\bibitem{Matschull} H.~J.~Matschull,
{On the relation between (2+1) Einstein gravity and Chern-Simons Theory}
{ Class.~Quant.~Grav.}  16 (1999) 2599--2609.
\bibitem{MS5}C.~Meusburger and B~J.~Schroers,  Phase space structure of Chern-Simons theory with a non-standard puncture, Nucl.~Phys.~B 738 (2006) 425--456;  hep-th/0505143.
\bibitem{tHooft}
  G.~'t Hooft,  Quantisation of point particles in 2+1 dimensional gravity and space-time
  discreteness,   Class.~Quant.~Grav.~13 (1996) 1023-1039;  arXiv:gr-qc/9601014.
\bibitem{MatWell} H.~J.~Matschull and M.~Welling, 
Quantum mechanics of a point particle in (2+1)-dimensional gravity,
Class.~Quant.~Grav.~15 (1998) 2981--3030; arXiv:gr-qc/9708054.
\bibitem{AB}
M.~Atiyah and R.~Bott, Yang-Mills equations over Riemann surfaces,
Phil.~Trans.~R. ~Soc.~London, {308} (1982) 523--615.
\bibitem{Atiyah}M.~Atiyah, The geometry and physics of knots,
  Cambridge University Press, Cambridge 1990.
 \bibitem{Schroerstalk} B.~J.~Schroers,   Lessons from (2+1)-dimensional quantum gravity,
 PoS QG-PH:035, 2007; 0710.5844 [gr-qc].
\bibitem{AMII} A.~Y.~Alekseev and A.~Z.~Malkin, Symplectic
structure of the moduli space of flat connections on a Riemann
surface', Commun.~Math.~Phys.{ 169} (1995)  99-119.
\bibitem{CP} V.~Chari and  A.~Pressley, { Quantum Groups, } Cambridge University Press,  Cambridge, {1994}.
\bibitem{Mackey}
G.~W. Mackey, Imprimitivity for representations of locally compact groups.
  {I},  Proc.~Nat.~Acad.~Sci. U. S. A.  35 (1949) 537--545.
  \bibitem{BaRa} A.~O.~Barut and  R.~Raczka, {Theory of group
representations and applications}, World Scientific, Singapore,1986.
\bibitem{KK} E.~Kiritsis and C.~Kounnas,  Phys.~Lett.~B320 (1994) 264--272, Addendum ibid B 325 (1994) 536
\bibitem{KM} T.~Koornwinder and  N.~Muller, The quantum double
of a (locally) compact group, {Journal of Lie Theory} {7} (1997)
101--120.
\bibitem{BM} F.~A.~Bais and  N.~Muller,
Topological field theory and the quantum double of
  $SU(2)$, { Nucl.~Phys.} { B530} (1998) 349-400.
  \bibitem{BDWP} F.~A.~Bais, P.~van~Driel and  M.~de~Wild~Propitius,  Quantum symmetries in discrete gauge theories, Phys.~Lett.~B280 (1992) 63--70; hep-th/9203046.
\bibitem{KMB} T.~Koornwinder and N.~Muller, F.~A.~Bais, Tensor
product representations of the quantum double of a compact group,
{ Commun.~Math.~Phys} {198} (1998) 157-186.
\bibitem{Gott} J.~R.~Gott, Closed timelike curves produced by
pairs of moving cosmic strings: exact solutions, Phys.~Rev.~Lett.
 66 (1991) 1126--1129.
\bibitem{MS3} C.~Meusburger and B~J.~Schroers,
 Mapping class group actions in Chern-Simons theory with gauge group $G\ltimes {\mathfrak g}^*$, Nucl.~Phys.~B 706 (2005) 569--597; hep-th/0312049.
\bibitem{tHooft2} G. 't Hooft, Non-perturbative 2 particle scattering
amplitude in 2+1 dimensional quantum gravity, Commun. Math. Phys. { 117}
(1988) 685--700.
\bibitem{DJ} S. Deser and R. Jackiw,  Classical and quantum scattering
on a cone, Commun. Math. Phys. { 118} (1988) 495--509.
\bibitem{BaMa} E.~Batista and  S.~Majid,
 Noncommutative geometry of angular momentum space U(su(2)),  J.~Math.~Phys.~44 (2003) 107-137;  hep-th/0205128.
 \bibitem{FM}L.~Freidel  and S.~Majid, 
  Noncommutative harmonic analysis, sampling theory and the Duflo map in 2+1 quantum gravity,  Class.~Quant.~Grav.25 (2008) 045006; hep-th/0601004. 
  \bibitem{MS} B.~J.~Schroers and S.~Majid,
   q-Deformation and Semidualisation in 3d Quantum Gravity, arXiv:0806.2587.


%%%%%%%%%%%%%PUT INTO RIGHT PLACE%%%%%%%%%



%%%%%%%%%%%%%%%%%%%%%%%%%%%%%%%%%
% following are not referred to yet
%%%%%%%%%%%%%%%%%%%%%%%
%

% \bibitem{MS7}C.~Meusburger and B.~J.~Schroers, Generalised Chern-Simons actions for 3d gravity and $\kappa$-Poincar\'e symmetry, Nucl.~Phys.~B 806 (2009) 462--488, 2009; 
%arXiv:0805.3318.

%\bibitem{BR} E.~Buffenoir, P.~Roche, Harmonic analysis on
%the quantum Lorentz group,
%{Commun.~Math.~Phys.} {207} (1999)  499-555.

%

%\bibitem{BBH} A.~Ballesteros, N.~Rossano Bruno and F.~J.~Herranz, Non-commutative relativistic
%spacetimes and worldlines from 2+1 quantum (anti) de Sitter groups, hep-th/0401244.

%%\bibitem{KM} P.~Kosinski and P.~Maslanka, 
%%The $\kappa$-Weyl group and its algebra, q-alg/9512018
%%\bibitem{KMLS} P.~Kosinski, P.~Maslanka,J.~Lukierski and A.Sitarz, Generalised $\kappa$-deformations and 
%%deformed relativistic scalar fields on noncommutative Minkowski space, hep-th/0307038.
%\bibitem{BBH1}A.~Ballesteros, N.~Rossano Bruno, and F.~J.~Herranz, 
%On 3+1 anti-de Sitter and de Sitter Lie bialgebras with dimensionful deformation parameters,
%hep-th/0408196

%
%   \bibitem{DH2} C.~Duval  and  P.~A.~Horvathy,  Chern-Simons gravity, based on a non-semisimple group, arXiv:0807.0977 [hep-th] 
% %this paper talks about CS theory of NW
% \bibitem{MaRa} J.~E.~Marsden, T.~S.~Ratiu, {Introduction to
%mechanics and symmetry}, Springer Verlag, New York, 1999.

%
% \bibitem{Landsman} N.~P.~Landsman,
%Mathematical topics between  classical and quantum mechanics,
%Springer Verlag, New York, 1998.


\end{thebibliography}

\end{document}